\documentclass[aps,reprint,onecolumn,nofootinbib]{revtex4-1}
\usepackage{graphicx,amsfonts,amsmath,amssymb}
\usepackage{bm}
\usepackage{url}
\makeatletter
\g@addto@macro{\UrlBreaks}{\UrlOrds}
\makeatother

\begin{document}
\title{\mint: A Fast Lightweight Envelope/Monte-Carlo Beam Optics
  Code for the Proton Beamlines of the Paul Scherrer Institute}
\date{\today}
\author{C. Baumgarten}
\affiliation{Paul Scherrer Institute, Switzerland}
\email{christian.baumgarten@psi.ch}

\def\begeq{\begin{equation}}
\def\endeq{\end{equation}}
\def\begary{\begeq\begin{array}}
\def\endary{\end{array}\endeq}
\def\bmtx{\left(\begin{array}}
\def\emtx{\end{array}\right)}
\newcommand{\myarray}[1]{\begin{equation}\begin{split}#1\end{split}\end{equation}}
\def\eps{\varepsilon}
\def\g{\gamma}
\def\y{\gamma}
\def\w{\omega}
\def\W{\Omega}
\def\s{\sigma}
\def\mint{\textsc{MinT}}
\def\opal{\textsc{OPAL}}
\def\beginfig{\onecolumngrid}
\def\endfig{\onecolumngrid}
\def\Exp#1{\exp\left(#1\right)}
\def\Log#1{\ln\left(#1\right)}
\def\Sinh#1{\sinh\left(#1\right)}
\def\Sin#1{\sin\left(#1\right)}
\def\Tanh#1{\tanh\left(#1\right)}
\def\Tan#1{\tan\left(#1\right)}
\def\Cos#1{\cos\left(#1\right)}
\def\Cosh#1{\cosh\left(#1\right)}

\begin{abstract}
  We report about the methods used in, and the performance of, the new fast
  and light-weight linear beam transport program \mint. \mint\ provides,
  beyond the usual linear ion optics, methods to compute the effects of beam
  degradation, multiple scattering and beam collimation. 
  This is specifically important in facilities where the ion beam passes matter, 
  for instance in proton therapy beamlines with an energy degrader as in the
  Proscan facility at PSI, but also for modeling the beam traversing the
  Muon- and Pion-production targets of the Paul Scherrer Institut's high
  intensity proton accelerator (HIPA). \mint\ is intended to be useful as a
  support tool for the HIPA and Proscan control rooms. This requires to
  have useful results within a few seconds. Hence simplicity and speed
  of calculation is favored against higher accuracy. 

  \mint\ has been designed not only to replace the FORTRAN 77 codes
  TRANSPORT and TURTLE, but to combine and extent their capabilities.
  \mint\ is a byte-code-compiler which translates an input language,
  described by syntactic rules. This allows for control structures
  like ``if-then-else'' or ``while''-loops, thus providing a high
  flexibility and readability. 
\end{abstract}

\pacs{45.50.Dd,87.56.bd,28.65.+a}
\keywords{Particle Accelerators, Accelerators in Radiation therapy}
\maketitle

\section{Introduction}

The Paul Scherrer Institute (PSI) in Villigen, Switzerland, is known for
it's high intensity proton accelerator (HIPA)~\cite{HIPA07,HIPA10,hipaSciPost}
which held for many years the world record of proton beam power (up to
$1.4\,\rm{MW}$). PSI is also known for pioneering research of proton tumor therapy in the
center for proton therapy (CPT) and it's facility Proscan~\cite{ZPT1}. 

Both facilities are driven by cyclotrons and in both facilities the 
simulation of the proton beam optics faces the problem to describe
beams of several ten to some hundred $\rm{MeV}$ passing through matter 
with subsequent beam collimation: In the HIPA facility, the $590\,\rm{MeV}$
proton beam has to pass through two graphite wheel targets before it is
send to the SINQ spallation neutron source. In case of Proscan, the beam
passes through an adjustable graphite wedge degrader to tune the beam
energy and hence the range of the protons in the target tissue.

\begin{figure}[b]
\parbox{170mm}{
\includegraphics[width=170mm]{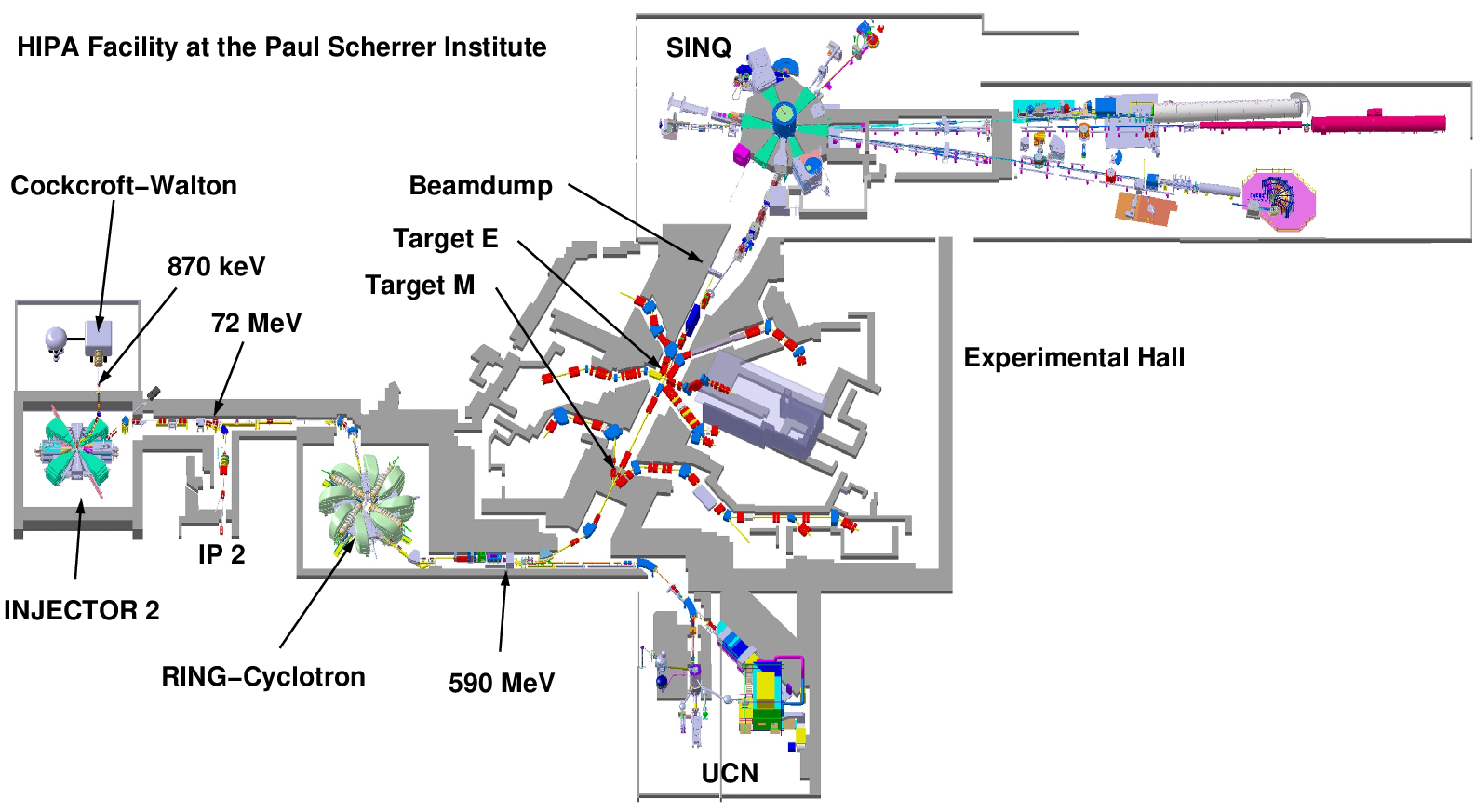}
\caption[HIPA Overview]{
Overview over HIPA facility at PSI. Here we are concerned specifically with
the beamlines, the $870\,\rm{keV}$ injection beamline ``BW870'', the
$72\,\rm{MeV}$-beamline ``IW2'', connecting Injector 2 with the Ring
cyclotron and the $590\,\rm{MeV}$ beamline, the so-called ``p-Channel'',
which guides the proton beam from the Ring cyclotron via Target M and Target E
to beamdump and SINQ, respectively.   
\label{fig_hipa}}
}
\end{figure}

The HIPA facility, shown in Fig.~\ref{fig_hipa} is driven by an accelerator
chain consisting of a Cockcroft-Walton-type pre-accelerator and two
isochronous separate sector cyclotrons, namely the $72\,\rm{MeV}$ Injector II
and the Ring cyclotron, providing a $590\,\rm{MeV}$ proton beam of up to
$2.4\,\rm{mA}$. This high intensity beam is either send to the
ultra-cold neutron source ``UCN''~\cite{UCN0,UCN1,UCN2,UCN3,UCN4,UCN5,UCN6}
or via two graphite targets, the $5\,\rm{mm}$ thick {\it Target M} and the
$40-60\,\rm{mm}$ thick {\it Target E}~\cite{Targets0,Targets1,hipaTargets},
to the Swiss spallation neutron source ``SINQ''~\cite{SINQ0,SINQ1,SINQ2,SINQ3,SINQ4,SINQ5}. 
\beginfig
\endfig

The Proscan therapy (PT) facility is driven by a $250\,\rm{MeV}$ superconducting
compact cyclotron ``COMET'' with currents of (usually) up to $1\,\rm{\mu
  A}$~\cite{Blosser93,Schillo2001,Geisler2004,Klein2005,Geisler2007,COMET1,COMET2,COMET3,COMET4}. 
After extraction from COMET, the $250\,\rm{MeV}$ beam passes a
double-wedge degrader which is used to reduce the beam energy to
$74-230\,\rm{MeV}$. The energy degradation increases the emittance and
energy spread by multiple Coulomb scattering. After energy degradation,
the beam has to be collimated and the energy spread reduced in order
to match the acceptance of the beam transport system. Furthermore, the
degraded beam passes several thin monitors and vacuum windows before 
entering a treatment room.

In any accelerator facilities it is highly desired to have a
fast beam-optics software to support control room work, like the
prediction of beam envelopes, the fit of envelopes to measured
profiles or the development of new tunes. A Beam dynamics code
that is supposed to be useful for both, the HIPA and Proscan
beamlines, must hence be able to model the effects that degraders
and collimators have on beam parameters like energy, emittance
and beam divergence.
Up until recently, the only online beam dynamics code used in
the HIPA control room was Urs Rohrer's graphical version of
TRANSPORT~\cite{Rohrer,transport1,transport2}. The Proscan control rooms
had no beam dynamics tool at all, since the use of TRANSPORT and
TURTLE~\cite{turtle1,turtle2}, though possible in principle, turned out
to be too cumbersome to support daily work in the control rooms.
High-accuracy codes like \opal\cite{opal2009a,opal2009b,opal1,opal2,opal2019}, 
GEANT~\cite{GEANT1,GEANT2,GEANT3}, or the Geant-based BDSIM~\cite{BDSIM},
do exist, but are better suited for high-accuracy offline studies,
preferably executed on high performance computers.

\mint\footnote{(M)int (i)s (n)ot (T)ransport~\cite{MintMan}.} is a lightweight
beam optics code designed to support control room activities,
for instance for beam tuning and development, that allows to provide reasonably 
accurate answers in short times (a few seconds up to a minute) instead of 
highly accurate results within tens of minutes or hours. Though \mint\ 
has been specifically designed as an online tool, it is useful offline as
well, as a design-tool for the layout of new beam lines, when the speed of
calculation and the flexibility of the code are essential as well. 
Since modern control room computers are mostly using Linux as operating
system, \mint\ is a Linux-program as well. Other operating systems are
not (yet) supported. However, \mint\ is fast enough for the use in
virtual machines. 

In the following we report about the features and performance of \mint,
the methods used in the program and demonstrate it's capabilities
to model the beam optics of the beamlines which are part of PSI's proton
facilities.

\section{The Beam Optics Program \mint}

A precursor of \mint\ was developed for the Gantry 3 project at 
PSI~\cite{G3a,G3b,G3c,Rizzoglio,G3ecpm}. The main idea was to model
the transition through matter (i.e. energy degrader) by fast 
and simple approximations and through collimators by a
``removal on hit'' strategy with a Monte-Carlo-ensemble created 
from the matrix of second moments. This is a pragmatic ``engineering''
kind of approach which often is sufficient in accuracy with respect to
the transmitted beam. \mint\ is a complete re-write of this first
code, which is now controlled by a programming style input file
with capabilities for graphical output to screen and to all file formats
supported by the GNU plotutils library (notably Postscript, PNG and FIG).

\subsection{Optics Calculations in \mint}

The ion optics machinery of \mint\ is based on the linear beam optics
methods known from the programs TRANSPORT and TURTLE, i.e. the beam is
described by it's first and second moments in local co-moving coordinates.
If $X=(x,x',y,y',z,\delta)^T$ are the usual six phase space coordinates
in the co-moving frame~\footnote{
Where the dash indicates the derivative along the beam path 
$x'\equiv {dx\over ds}$ and $\delta={dp\over p}$ is the momentum spread.}, 
then the first moments are $X_i(s)=\langle X_i\rangle_s$
and second moments are given by a symmetric matrix 
$\Sigma_{ij}(s)=\langle (X_i-\langle X_i\rangle)\,(X_j-\langle X_j\rangle)\rangle_s$. 
The particle transport through beam optical elements, using a
linear~\footnote{\mint, starting with version 0.50, supports second 
order calculation. But since second order effects are weak in the PSI 
proton beamlines, only first order has been compared and tested against 
measurements.} approximation and hard-edge magnets, is then 
given by a sequence of multiplications with symplectic matrices ${\bf M}_k$.
\mint\ allows to control the stepsize in all beamline elements individually
and can therefore be optimized by the user either for speed or for accuracy. 
It is possible, but not always desired, to compute the transfer matrix of
some element in one step, since beam-loss typically occurs inside of quads and
bends. Possible beam-loss in the center of magnets can only be estimated 
with reasonable accuracy, if the beam optics inside the element is calculated 
in small steps. \mint\ evaluates beam losses only when the envelope is 
evaluated as well (i.e. {\it after} some step). \mint\ posesses two modes,
a pure envelope mode, corresponding to linear TRANSPORT calculations, and a 
Monte-Carlo assisted envelope mode (``sampled mode''), where the beam is
represented by a Monte-Carlo generated ensemble of ``rays'' $X_i$. The
user can switch between these modes by the insertion of dedicated elements.

In envelope mode the quations are:
\myarray{
X(s+L)&={\bf M}_k(L)\,X(s)\\
\Sigma(s+L)&={\bf M}_k(L)\,\Sigma(s)\,{\bf M}_k(L)^T\\
}
where $s$ is the position along the reference trajectory, $L$ is the length of
the calculation step in element $k$ and ${\bf M}_k$ is the transfer matrix of
the k-th element.  
\mint\ is equipped with an integrated symplectic Monte Carlo generator for 
multivariate Gaussian distributions~\cite{stat_paper}. A distribution of $n$ 
trajectories is represented by a $6\times n$-matrix ${\bf X}_{ij}$, such that 
the center of the bunch is given by
\begeq
X_i(s)=\frac{1}{n}\,\sum\limits_{j=1}^n\,{\bf X}_{ij}(s)
\endeq
and the centered distribution ${\bf\tilde X}_{ij}$ by
\begeq
{\bf\tilde X}_{ij}={\bf X}_{ij}(s)-X_i(s)\,.
\endeq
The matrix of second moments is then
\begeq
\Sigma_{ij}=\frac{1}{n}\,{\bf\tilde X}\,{\bf\tilde X}^T
\endeq
Without collimators, the evolution of the beam centroid, MC-sample and 
matrix of second moments is described by a symplectic transfer
matrix ${\bf M}(L)$ and is in Monte-Carlo mode given by:
\myarray{
{\bf X}(s+L)&={\bf M}_k(L)\,{\bf X}(s)\\
\Sigma(s+L)&={\bf\tilde X}(s+L)\,{\bf\tilde X}(s+L)^T
}
In Monte-Carlo-mode, the $\Sigma$-matrix is obtained
from the (possibly re-centered) Monte-Carlo-sample.

In the presence of collimators, the MC-sample is ``filtered'':
trajectories which hit a collimator, are  removed from the
ensemble ${\bf X}$ which represents the beam. The current version
of \mint\ allows for circular, elliptic and rectangular collimators
and moveable horizontal or vertical, symmetric or asymmetric, slits.
If $X=(x,x',y,y',z,\delta={dp\over p})^T$ is the vector of local
coordinates, then a trajectory passes an elliptic collimator 
with horizontal and vertical half-diameter $a$ and $b$, if
\begeq
{(x+x_0)^2\over a}+{(y+y_0)^2\over b}\le 1
\label{eq_test}
\endeq
where $x_0$ and $y_0$ are the coordinates of the beam centroid.

No scattering calculations are done on collimators. But in many
beam optical calculations, it is not the primary objective to 
obtain a detailed calculation of losses and activation, but
the knowledge of the properties of the fraction of the beam which
passes the collimator. The test Eq.~\ref{eq_test}
of $10^4-10^6$ trajectories is computationally expensive, but  
the simplest general scalable method to solve the (strongly non-linear)
collimation problem.

\subsection{Beam Degradation}

Besides collimation, the main task appearing for instance in 
proton therapy beamlines with beam degrader is the calculation of
beam degradation effects, such as energy reduction and -straggling, and emittance 
increase by multiple scattering~\cite{Scharf,ChaoChou,Newhauser,Durante,Paganetti}.

\subsubsection{Energy Degradation}

For the energy loss (and range) calculations the Bethe-Bloch-formula
is used to describe the average energy loss per unit length of charged
particles passing through matter~\cite{Groom}:
\begeq
{dE\over ds}=-K\,z^2\,\frac{Z}{A}\,\frac{1}{\beta^2}\,\left[\frac{1}{2}\,\log{({2\,m_e\,c^2\,\beta^2\,\gamma^2\,T_{max}\over I^2})}-\beta^2-\frac{\delta}{2}\,\right]\,.
\label{eq_bethebloch}
\endeq
where $s$ is the pathlength, $m_e$ is the electron mass, $z$ and $M$ are the
charge number and the mass of the projectile, $N_A$ is Avogadro's constant,
$Z$ and $A$ are atomic number and mass of the medium (in $g\,mol^{-1}$),
and $K$ is the factor $K=4\,\pi\,N_A\,r_e^2\,m_e\,c^2$.
Furthermore $r_e$ is the classical electron radius $r_e={e^2\over
  4\,\pi\,\eps_0\,c^2}$, $\beta=\frac{v}{c}$ is the velocity in units
of the speed of light $c$, and $\gamma$ is the known relativistic factor
$\gamma={1\over\sqrt{1-\beta^2}}$. $I$ is the mean excitation energy in 
units of $eV$, $\delta$ is the density effect correction to ionization energy
loss, and $T_{max}$ is finally given by 
\begeq
T_{max}={2\,m_e\,c^2\,\beta^2\,\gamma^2\over 1+2\,\gamma\,m_e/M+(m_e/M)^2}\,.
\endeq
The current version of \mint\ neglects the {\it density effect correction term} 
(i.e. $\delta\to 0$) as the contribution is negligible in the energy range
below about a $GeV$~\cite{Groom}.

In case of degraders made from composite materials (containing different
nuclei), \mint\ computes the energy losses for each target component with the 
corresponding reduced density:
\begeq
{dE\over ds}=-\frac{K\,z^2}{\beta^2}\,\sum\limits_{k}\,\frac{Z_k}{A_k}\,\rho_k\,\left[\log{({2\,m_e\,c^2\,\beta^2\,\gamma^2\over I_k})}-\beta^2\right]\,,
\label{eq_bbc}
\endeq
where the {\it partial} densities $\rho_k$ have to sum up to the density of the compound $\rho$:
\begeq
\rho=\sum\limits_{k}\,\rho_k\,.
\endeq
For the convenience of use, \mint\ provides the possibility define
arbitrary pure and composite materials.

\subsubsection{Energy Straggling}

The conventional 6th beam coordinate $\delta={\delta p\over p_0}$ is a
relative quantity and with the degradation process, both the momentum spread
$\delta p$ and the average momentum $p_0$, change. The former increases by
energy straggling while the latter decreases by energy degradation.

The latter means that the final energy $E_f$ depends, after the passage
through a slab of fixed thickness $\Delta s$, non-linearily on the initial 
energy $E_i$, so that
\begeq
{dE_f\over dE_i}={{dE_f/ds}\over{dE_i/ds}}
\endeq
and hence
\begeq
\sigma_{E_f}^2=\left({{dE_f/ds}\over{dE_i/ds}}\right)^2\,\sigma_{E_i}^2\,.
\endeq
The combination with the stochastic straggling yields~\cite{Bednyakov}:
\begeq
\sigma_{E_f}^2=\left({M(E_f)\over M(E_i)}\right)^2\,\left(\sigma_{E_i}^2+M(E_i)^2\,\int\limits_{E_f}^{E_i}\,{N(T)\over M(T)^3}\,dT\right)\,,
\label{eq_bednyakov}
\endeq
where $M(E)=-{dE\over ds}$ and $N(E)$ is the stochastic energy straggling for
thin targets (in dependence on energy), which can be calculated by~\cite{Leo}:
\myarray{
N(E)&={d\sigma_E^2\over ds}=4\,\pi\,N_a\,r_e^2\,(m_e\,c^2)^2\,\rho\,{Z\over
  A}\,(1-\frac{1}{2}\,\beta^2)\,\gamma^2\\
&=4\,\pi\,N_a\,r_e^2\,(m_e\,c^2)^2\,\rho\,{Z\over A}\,{1+\gamma^2\over 2}\\
}
where $\rho$ is the density, $Z$ the nuclear charge and $A$ the mass number of the scatterer.
If we change the integration variable in Eqn.~\ref{eq_bednyakov}, we obtain in the limiting case $\Delta x\to 0$:
\myarray{
{\sigma_{E_f}^2\over M(E_f)^2}&={\sigma_{E_i}^2\over M(E_i)^2}+\int\limits_0^{\Delta x}\,{N(E(x))\over M(E(x))^2}\,dx\\
{d\over ds}\left({\sigma_{E}^2\over M(E)^2}\right)&={N(E)\over M(E)^2}\\
\label{eq_bednyakov2}
}
It is hence necessary to normalize the energy variance with the (absolute value of) the Bethe-Bloch function $M(E)$ 
in order to obtain, in Gaussian approximation, the desired differential
equation. Eq.~\ref{eq_bednyakov2} can be integrated numerically to finally
obtain $\sigma_E^2$ at the final energy $E_f$ and hence the final momentum
spread $\delta$. 

\subsubsection{Deep Inelastic Scattering}

\mint\ enables to estimate the beam loss by deep inelastic scattering (DIS) by
the formula provided in Ref.~\cite{DIScat1,DIScat2,DIScat3} for energies between a few
$\rm{MeV}$ and a few hundred $\rm{MeV}$. Any projectile that is subject to
some inelastic process, is counted as lost. But \mint\ does not remove tracks
due to DIS, but simply calculates the surviving fraction of the beam. The beam
current is then reduced along the beam path accordingly.

The DIS calculation requires that the user specifies the RMS radius $R_{rms}$
for the projectile and passed material. Then an inverse scattering length 
${1\over\lambda_{DIS}}$ can be written as
\begeq
{1\over\lambda_{DIS}}={\rho\over A}\,\sigma_{DIS}
\endeq
The beam current change $dI$ for passing a slab of thickness $dx$ is
then approximated by
\begeq
{dI\over ds}=-{I\over\lambda_{DIS}}
\endeq
Hence the surviving fraction of particles without deep inelastic process 
decreases exponentially with the thickness of the traversed matter.

\subsubsection{Emittance Increase by Lateral Straggling}

\mint provides two scattering models, the first follows the suggestion
of Francis Farley~\cite{farley1,farley2}, which is a ``local'' approximation
of the Moliere theory. Farley refers to Ref.~(\cite{Groom}) where the
precision is claimed to be $11\,\%$. The theoretical description of the beam
passage through the solid degrader material can best be understood, if one
considers an idealized parallel beam (zero emittance)
interacting with a single scatterer. The immediate effect of the scatterer is to
change {\it only angle} (and energy) of the incident proton (or ion). 
The time derivative of the matrix $\sigma$ of second moments is then
given (in one dimension) by~\cite{farley2}:
\begeq
{d\sigma\over ds}=\bmtx{cc}2\,\sigma_{12}&\sigma_{22}\\\sigma_{22}& T(p)\emtx\,,
\label{eq_eqom_moments}
\endeq
where $T(p)$ is the scattering power. Farley used $T_{FF}(p)={K\over
  p^2\,\beta^2}$ as scattering power where $p=p(s)$ is the momentum, 
$\beta=\beta(s)=v(s)/c$ is the ratio of the projectile's velocity to 
the velocity of light and $K$ is a material dependent constant that 
is given by Farley as:
\begeq
K=200\,(\rm{MeV/c})^2\,{Z^2\over X_0}\,,
\endeq 
with the charge of the projectile $Z$ and the radiation length of
the target material $X_0$~\cite{Groom} given in units of $\mathrm{[g\,cm^{-2}]}$:
\begeq
X_0={716.4\,A\over Z\,(Z+1)\,\log{(287/\sqrt{Z})}}\,.
\endeq
Often an alternative radiation length $\tilde X_0=X_0/\rho$ is used, which is
devided by the target density $\rho$ and then has the unit $\mathrm{[cm]}$.
\beginfig
\begin{figure}[t]
\parbox{17.0cm}{
\includegraphics[width=170mm]{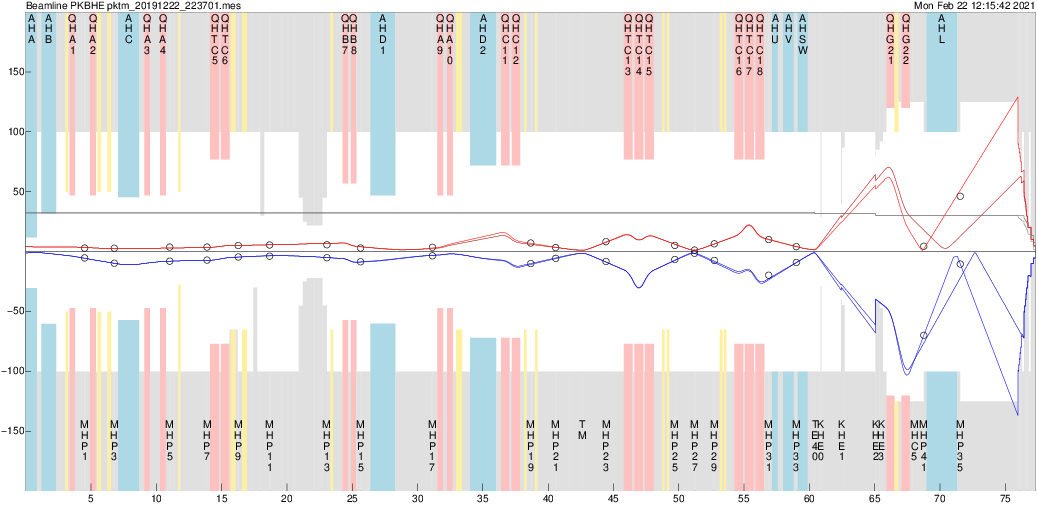}
\caption[Envelopes from Ring Cyclotron to PSI Beam Dump BHE]{
TRANSPORT style plot ($2\,\sigma_y$ in red, $-2\,\sigma_x$
in blue) of the beam envelopes of the HIPA proton channel 
starting from Ring cyclotron to the high energy beamdump 
``BHE'',
with a $60\,\rm{mm}$ ($40\,\rm{mm}$) long {\it Target E} in dark 
(light) blue/red. The symbols show the $2\,\sigma$-beam 
width as measured with beam profile monitors for the 
$40\,\rm{mm}$ long {\it Target E}. 
The $60\,\rm{mm}$ data have been taken with a more focused 
beam on beam dump. The (calculated) surviving beam currents
are shown logarithmically in dark gray (i.e. $10\,\log{(I/\rm{\mu A})}$).
\label{fig_pkbhe}}}
\end{figure}
\endfig

According to Gottschalk, this corresponds to the scattering power of 
Fermi and Rossi~\cite{Gottschalk2}:
\begeq
T_{FR}=\left({E_s\,Z\over p\,\beta}\right)^2\,{1\over X_0}\,,
\endeq
where $E_s=15\,\mathrm{MeV}$ approximately matches Farley's formula. 
Gottschalk presented a detailed comparison of various theoretical models 
in Ref.~(\cite{Gottschalk2}) and derived a phenomenological model to match 
the experimental data. His scattering power is
\begeq
T_{GS}=f_{dM}\,\left({E_s\,Z\over p\,\beta}\right)^2\,{1\over X_S}\,,
\label{eq_theta_gs}
\endeq
where he replaced the radiation length $X_0$ by the ``scattering length'' $X_S$, defined by
\begeq
X_S={\alpha_r\,A\over N\,Z^2\,r_e^2\,\rho}\,{1\over 2\,\ln{(33219\,(A\,Z^{-1/3}))}-1}\,,
\endeq
and the weight function $f_{dM}$ is given by~\cite{Gottschalk2}
\myarray{
l_1&=\log{(p_{\mathrm{[MeV/c]}}\,\beta)}\\
l_2&=\log{(1+\eps-({p\,\beta\over p_1\,\beta_1})^2)}\\
f_{dM}&=0.5244+0.085773\,l_2+0.1\,l_1-0.004256\,l_1\,l_2\,,
\label{eq_GSf}
}
where $p_1$ and $\beta_1$ are momentum and velocity (in units of $c$) of the
incident projectile. Gottschalk's original formula is obtained for $\eps=0$.
However, for very thin sheets of material, Gottschalk's formula may result
in large negative values for $l_2$. The parameter $\eps$ enables the use
to prevent unphysical results for these cases.
The user may therefore use either Farley-Fermi-Rossi or Gottschalk's scattering 
power. A global variable named ``MSFactor'', defined by the user, allows to adjust
the strength of angular straggling.
The variable $f_{dM}$ in Eqs.~\ref{eq_theta_gs} and ~\ref{eq_GSf} are then
multiplied by this factor, if it has been defined. Otherwise it is omitted.

In any case, \mint\ integrates Eq.~\ref{eq_eqom_moments} to model the influence
of multiple Coulomb scattering on the $\Sigma$-matrix of the beam. If \mint\ 
runs in Monte-Carlo-mode at the entrance of some degrader element, then the
calculation is done twice in parallel: The existing ensemble is transported as
in case of a drift and an initially zero $\Sigma$-matrix (a zero-emittance beam) is
integrated according to Eq.~\ref{eq_eqom_moments} and then used to generate a 
second (scattered) ensemble. At the end of the element, the matrices of both 
ensembles are added, which means that the direction of each trajectory is 
changed statistically.

\section{Examples: \mint\ for PSI beamlines}
\label{sec_psi}

In this section we provide some example applications of \mint,
namely to the PSI proton beamlines.

\subsection{HIPA Beamlines}
\label{sec_hipa}

Fig.~\ref{fig_hipa} gives an overview of the HIPA
facilities~\cite{hipaSciPost}. The beam is
generated in a compact ECR proton source~\cite{ecr} and extracted with
a $60\,\rm{kV}$ extraction system which is located on the high
voltage platform of the $810\,\rm{kV}$ Cockcroft-Walton accelerator.
The resulting $870\,\rm{keV}$ DC beam with a current of $10-12\,\rm{mA}$
is send through a system of two bunchers, a $50\,\rm{MHz}$ first and
a $150\,\rm{MHz}$ third-harmonic buncher, to the axial injection 
line of Injector II. Injector II is a $72\,\rm{MeV}$ high-current isochronous 
cyclotron with four separate
sectors~\cite{Adam75,Olivo79,Schryber81,Joho85,Markovits87,Stetson92}.
After extraction, the beam is send via the $72\,\rm{MeV}$-beamline ``IW2'' to the so-called
``Ring-cyclotron''~\cite{Willax63,Blaser70,Willax72,Willax73,Joho75,Adam75b,JA78} where it is 
accelerated to $590\,\rm{MeV}$. A fraction of several ten $\rm{\mu A}$ can be
split off from the $72\,\rm{MeV}$-beam and send to the isotope production facility IP2.
The $590\,\rm{MeV}$ beam is either transported to the ultra-cold neutrons at 
UCN (typically in ``pulses'' of a few seconds) or otherwise send to the 
muon/pion production graphite targets, first the $5\,\rm{mm}$ thick ``Target
M'' and then to the $40-60\,\rm{mm}$ long ``Target E''.
After an appropriate collimation, the remaining useable beam is either send to 
the SINQ-target for neutron production or otherwise to a high-intensity
beam dump ($I_{max}\approx 1.7\,\rm{mA}$). 

Since TRANSPORT offers no convenient possibility to simulate the passage 
of high-energy particles through matter, the description of this beamline
with TRANSPORT requires to split the optics calculation from Ring-cyclotron to 
SINQ (or beamdump) into at least $3$ sections, from Ring to {\it Target M}, from 
{\it Target M} to {\it Target E} and from {\it Target E} to beamdump or SINQ, 
respectively.
Each of this sections is then treated seperately and it is up to the user
to verify the overall consistency. 
The \mint\ monte-carlo-mode (activated here behind {\it Target M}) allows not only 
to compute the beam optics in one go, but also to predict beam currents and 
``realistic'' (non-Gaussian) beam profiles. 
In the following we provide some examples.

\subsubsection{From Ring Cyclotron to the Beamdump}
\label{sec_pkbhe}

Fig.~\ref{fig_pkbhe} shows the result of a \mint\ optical calculation
from Ring cyclotron to the beamdump for two versions of {\it Target E}, 
$60\,\rm{mm}$ and $40\,\rm{mm}$ long (at $s\approx 61\,\rm{m}$). 
The symbols are measurements of the latter. 
The apertures of the respective beamline components are indicated 
by a colored background in PSI convention, i.e. blue
for bends, red for quads, gray for drifts and yellow for
steering magnets.
The graphical output follows the TRANSPORT conventions, i.e. the 
vertical beam size ($2\,\sigma_{rms}$) is shown above axis (red) 
and the horizontal beam size (blue) below axis.
The symbols indicate the beam size as measured by beam profile monitors.

The primary beam intensity was ($60\,\rm{mm}$ target) $1840\,\rm{\mu A}$, 
the beam intensity after collimation (measured by current monitor MHC5) 
was $1022\,\mu\rm{A}$, the \mint\ prediction is $1090\,\mu\rm{A}$
using Gottschalk's scattering power ($1054\,\mu\rm{A}$ for the
modified version).
In case of the $40\,\rm{mm}$ target, the measured (simulated) 
current after {\it Target E} was $1120\,\rm{\mu A}$ ($1111\,\rm{\mu A}$), 
using the modified version of Gottschalk's scattering power.

These calculations used a Monte-Carlo sample of $50000$ ``macro''-particles 
(Monte-Carlo mode starting behind {\it Target E}) and took less than five 
seconds on an average laptop computer~\footnote{The speed of calculation 
can vary significantly, depending on the used step size. The calculations 
shown here included a space-charge kicks at least every $100\,\rm{mm}$.}. 
The prediction of the correct beam current depends on many
details so that the agreement with the measured currents is quite
satisfactory.

After the passage of {\it Target M}, the estimated beam energy is
calculated to be $586.6\,\rm{MeV}$, after the $40\,\rm{mm}$ 
($60\,\rm{mm}$) long {\it Target E} it is $569.3\,\rm{MeV}$ 
($560.6\,\rm{MeV}$). The emittance values assumed at the exit of
the Ring cyclotron are $\eps_x=0.154\,\rm{mm mrad}$ and 
$\eps_y=0.366\,\rm{mm mrad}$, after {\it Target M} they increase to
$\eps_x=0.468\,\rm{mm mrad}$ and $\eps_y=1.63\,\rm{mm mrad}$,
after {\it Target E} ($40\,\rm{mm}$) to $\eps_x=1.035\,\rm{mm mrad}$ 
and $\eps_y=5\,\rm{mm mrad}$. The \mint\ simulation results in
a beam loss of $1.1\,\%$ in {\it Target M} and of $8.24\,\%$ in {\it Target E},
which agrees well with the numbers given in Ref.~\cite{KHE2c}~\footnote{
More details on the collimators can be found in 
Refs.~\cite{KHE2a,KHE2b,KHE2c,KHE2d}.}.

\subsubsection{From Ring Cyclotron to SINQ}
\label{sec_pksinq}

\begin{figure}[t]
\parbox{17cm}{
\parbox{10.5cm}{
\includegraphics[width=100mm]{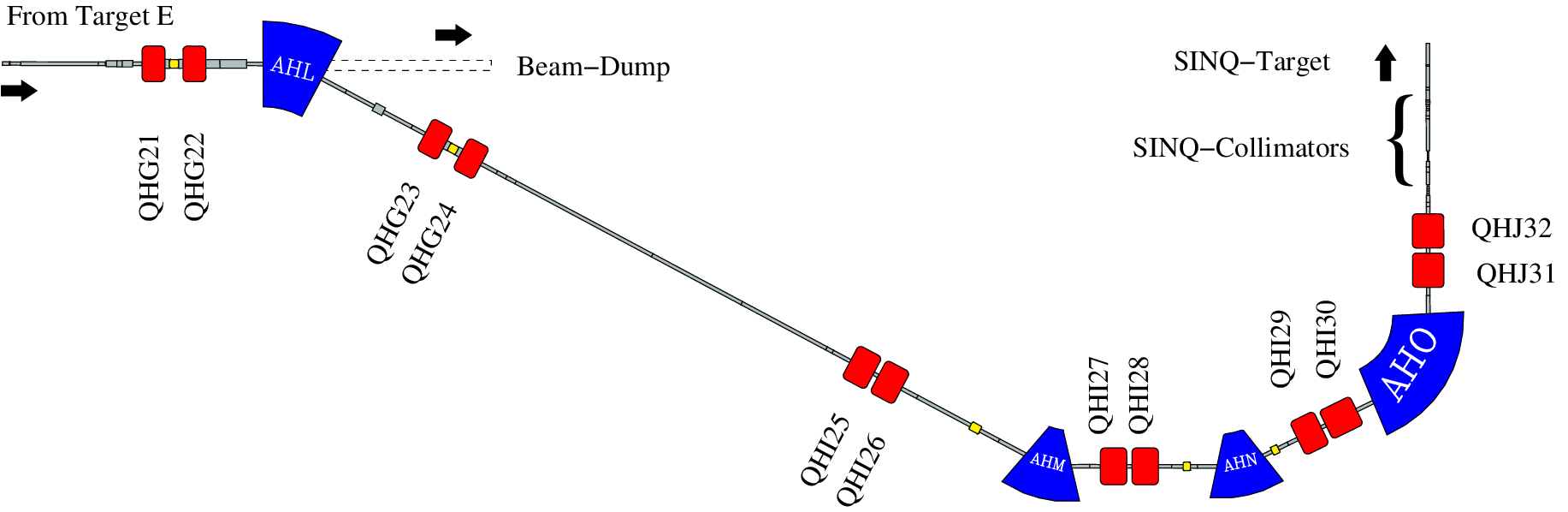}
}\hfill\parbox{6.5cm}{
\caption[]{
Beamline from switching-magnet AHL towards the spallation neutron 
source SINQ, shown as a side-view: the beam enters the SINQ target 
from below.
\label{fig_pkeller}}}}
\end{figure}
Since the installation of the SINQ, the beamdump is used only 
in times when the SINQ is not ready for data taking. In SINQ
operation, the magnet AHL is active and the SINQ-beamline guides
the beam downwards in order to inject it from below into the 
SINQ target (see Fig.~\ref{fig_pkeller})~\cite{KHE2d}.
The SINQ-beamline is equipped with some large aperture quadrupoles.
Rohrer's version of TRANSPORT supports large aperture quads by a
fringe field corrections derived from fringe field integrals~\cite{Rohrer} 
according to the approximation developed by Matsuda and
Wollnik~\cite{Matsuda}. However, the suggested corrections are not 
symplectic. Furthermore the calculation of the fringe field integrals 
requires a precise knowledge of the field shape, which is not always
available and/or reliable. 
Nonetheless \mint\ provides a symplectified version of this method,
described in Sec.~\ref{sec_qfringes}. Furthermore \mint\ allows to 
choose another (fully symplectic) correction due to Baartman which 
does not require the knowledge of the fringe field integrals. In case
of the SINQ beamline it provides an equivalently satisfying agreement
with the measured beam profiles.

If $f$ is the (uncorrected) focusing length of the quad $R$ the pole
radius and $L$ the effective length, then the Baartman's 
correction is based of the following change of the focusing 
length~\cite{Baartman}:
\begeq
\Delta f=R^2/(2\,L)
\endeq
A more detailed description of the implementation in \mint\ is 
given in Sec.~\ref{sec_qfringes} of the appendix.
\beginfig

\begin{figure}[t]
\parbox{17.0cm}{
\includegraphics[width=170mm]{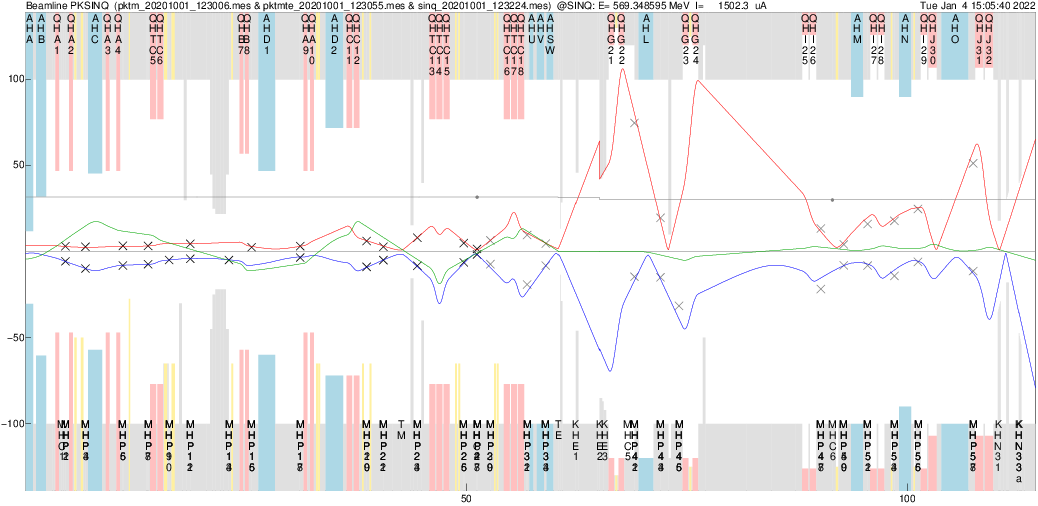}
\caption[Envelope from Ring Cyclotron to SINQ]{
Beam envelope from Ring cyclotron to SINQ target with the
$40\,\rm{mm}$ long {\it Target E} (Data from 2021 Production). 
The measured current towards SINQ was $955\,\rm{\mu A}$, 
the prediction of \mint\ is $952\,\rm{\mu A}$ with the modified 
version of Gottschalk's scattering power. The (scaled) dispersion 
is shown as a green solid line. The apertures of magnets and 
collimator are indicated by the respective background colors.
The gray line (symbols) represent the calculated (measured) beam
current.
\label{fig_pksinq40}}}
\end{figure}
\endfig
Both methods are able to describe the optics of the SINQ
beamline with satisfactory accuracy.
Fig.~\ref{fig_pksinq40} shows the results using a MC sample of
$10^4$ particles with Baartman's fringe field correction. The
profile measurements between Ring (start) and Target M (labeled ``TM'')
were used to match the starting conditions. The remaining beamline
(Target M up to SINQ) is a forward calculation. The execution time
on an average laptop (including fit) was about $6\,\rm{s}$.

Even though the predicted losses can deviate from the measurements
by a few percent, \mint\ allows for a reasonable online prediction
of the beam optics. In a high current facility like HIPA,
where even small losses, in the order of a permille, can overheat
and melt components, the accuracy of the \mint\ model is certainly
not sufficient to omit the fine tuning by operators.
However approximate beam tunes can be elaborated and fine-tuned
by a stepwise increase of the beam current to its production value.

However, for (the commissioning or tuning of) low intensity 
beamlines like the ones used for proton therapy machine, MinT provides 
sufficient numerical accuracy to compute tunes which require little
or no correction by manual fine-tuning.

\subsubsection{Injector II and the $72\,\rm{MeV}$-beamline}
\label{sec_inj2}

The $72\,\rm{MeV}$ beamline (``IW2'') connects the extraction
of injector II with the injection of the Ring cyclotrons~\cite{Bi2011}.

As reported elsewhere~\cite{Schryber95,Adam95,Yang2008,cyc_paper,cyc2013,Kolano,cyc2019}, the 
Injector II cyclotron is operated in the space-charge dominated regime. 
\mint\ provides the possibility to define a sequence of elements as 
a ``ring'' so that the matched beam matrix $\Sigma_m$ can be  derived from the
matching condition, for given emittances and current. The specific problem in
case of non-negligible space charge is the fact that the one-turn-transfer
matrix depends on the strength of the space charge and hence on the beam size.
If ${\bf M}={\bf M}(I,\eps_i)$ is 
the one-turn-transfer-matrix, depending on beam current $I$ and beam 
emittances $\eps_i$, then the beam is matched, iff
\begeq
{\bf S}={\bf M}\,{\bf S}\,{\bf M}^{-1}\,,
\endeq
where - with ${\bf J}$ as the symplectic unit matrix - the matrix
of second moments for the matched distribution is given as
$\Sigma_m={\bf S}\,{\bf J}$. \mint\ makes use of the general
decoupling/diagonalization methods described in
Refs.~\cite{rdm_paper,geo_paper,stat_paper,jacobi_paper},
which allow to determine ${\bf S}$ and hence $\Sigma$ for arbitrary
symplectic transport matrix ${\bf M}$ and given proper eigen-emittances.
However, here ${\bf M}$ depends itself on (elements of) the $\Sigma$-matrix,
for instance on (square of the rms) beam size, then the problem can not be
solved analytically and it is required to use an iterative
scheme~\cite{cyc_paper}. \mint\ is
equipped with such a scheme and, for the beam conditions and currents of
Injector 2, typically less than $10$ iterations are required to find
the matched beam. With low beam current or if space charge is ignored,
ideal isochronous cyclotrons do not provide any longitudinal focusing
and the matching is hence undefined in the longitudinal direction.

The matched beam is then used as a starting condition for a fit to the
measured profiles of the $72\,\rm{MeV}$ beamline which connects Injector II
and the Ring cyclotron. The \mint\ calculations confirmed that the matched
beam assumption provides excellent starting conditons which allow to fit
the beam envelope of the IW2-beamline in few steps. At high currents, the
horizontal beam envelope requires almost no adjustments to match the
measured beam sizes.

In linear approximation, the cyclotron specific space charge effect connect
only horizontal and longitudinal motion, while the vertical motion is not
affected. In other words, within Injector II, the vertical beam size
is not directly coupled 
to the horizontal and longitudinal motion and hence can be fitted to the 
measured beam sizes without strong influence on the so-called
``vortex motion''~\cite{cyc_paper,cyc2013,cyc2019}.

\begin{figure*}[t]
\parbox{17.0cm}{
\includegraphics[width=170mm]{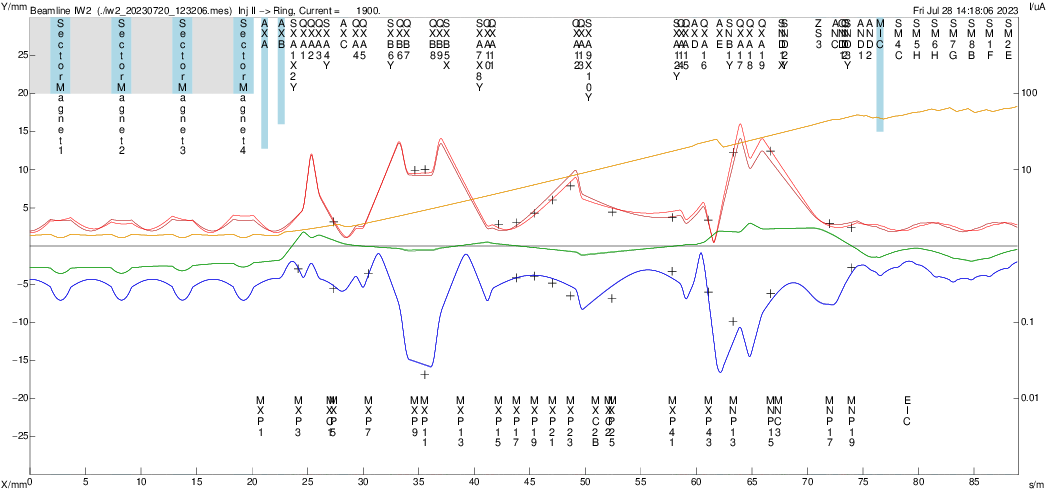}
\caption[Envelopes from Injector II to Ring Cyclotron]{
  Beam envelopes ($2\,\sigma$) from the last turn of 
  Injector II to the first turn of the PSI Ring cyclotron. 
  For the fit, it is assumed that the transverse horizontal 
  equals the longitudinal emittance ($\eps_{x,z}=\eps_x=\eps_z$). 
  The emittances $\eps_{x,z}$ and $\eps_y$ are fitted to the
  measured beam profiles, under the assumption that the
  beam is matched to the last turn of Injector II~\cite{cyc_paper,geo_paper}.
  The vertical (horizontal) envelopes are shown in dark red (blue).
  The ``+'' symbols indicate the respective beam sizes as
  measured by profile monitors and the light colored (red/blue) 
  lines indicate the results of \mint\ envelope calculation 
  before and after a fit of the axial beam parameters to fit
  the measured beam sizes. The corresponding initial (final)
  dispersion is shown in dark (light) green and the longitudinal
  beam dimension (scaled down by four) in (dark/light) orange.
  The computation, including all fits, takes a few seconds on 
  an average personal computer. In case of the IW2 beamline,
  where space charge has to be included, the computation speed
  depends strongly on the settings. In the plot shown here,
  the distance between two subsequent space charge kicks was 
  $50\,\rm{mm}$.
\label{fig_iw2_fit}}}
\end{figure*}

Fig.~\ref{fig_iw2_fit} shows the two stages of the fitting procedure.
The darker colors show the beam envelopes of a two-parameter fit 
(varying only two beam emittance values), assuming a matched beam 
from Injector II. In the second step, the vertical initial beam 
parameters ($\sigma_{33},\,\sigma_{34},\,\sigma_{44}$) are varied to 
improve the fit to the vertical beam sizes. On the left the simulation 
starts with the last turn of Injector II (four bends) and ends on the 
right with the first turn of the Ring cyclotron (eight bends). 
For a comparison with \opal\ see Ref.~\cite{Bi2011}.
\beginfig
\endfig

The results confirm that the horizontal and longitudinal beam
parameters are due to the space-charge induced coupling, while
the vertical beam parameters are linearily independent.
This kind of horizontal-longitudinal ``self-matching'' works,
as expected, only for sufficiently high beam currents. Their is no
passage through matter and no significant beam collimation in
the $72\,\rm{MeV}$ transfer line and \mint\ is therefore
operated here in pure envelope mode. 

\subsubsection{The $870\,\rm{keV}$-beamline}
\label{sec_bw870}

\begin{figure*}[t]
\parbox{17.0cm}{
  \includegraphics[width=170mm]{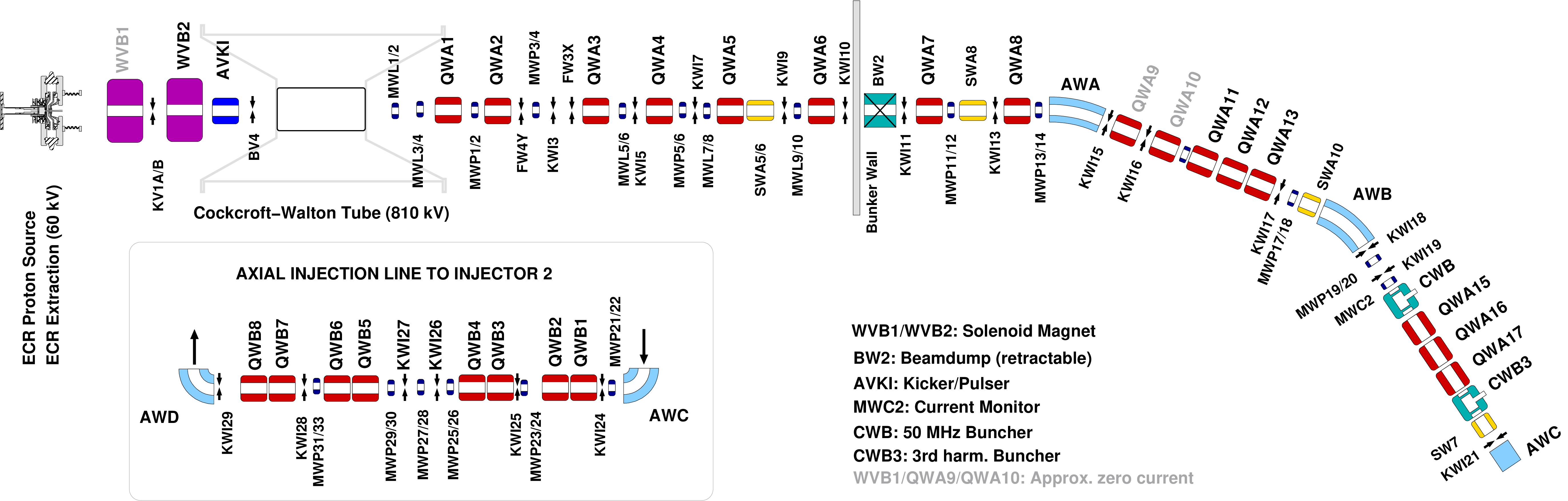}
}
\caption[Scheme of BW870]{
    Schematic overview of the HIPA beamline BW870 from ECR Ion source
    to Injector 2.
\label{fig_bw870}}
\end{figure*}

Fig.~\ref{fig_bw870} gives a schematic overview of the $870\,\rm{keV}$
``BW870'' injection line connecting the ECR ion source via Cockcroft-Walton DC
preaccelerator with the center of Injector 2~\cite{Olivo79,Markovits81,Olivo84,Olivo86,Markovits87}.
Due to the strong space charge forces of the $10\,\rm{mA}$ DC beam,
the beamline BW870 is most challenging in terms of ion-optical modelling.
The specific difficulty is due to (partial) space charge compensation by
electrons which are attracted and captured by the DC proton beam potential.
Furthermore, the DC proton beam is bunched by a first and third harmonic
buncher~\cite{Stetson92,CWB3_07}. The main task is here is to find a method
to model the transition from a DC beam into a bunched beam.
The bunching process involves all possible phases of the buncher and can
therefore not be modelled in linear approximation.

A rather simple method to model the effect of a buncher in combination with
a DC beam has been implemented in \mint. This method presumes a sampled beam.
Since the beam of the $870\,\rm{keV}$ beamline has passed a Cockcroft-Walton
type DC accelerator, the energy spread of the beam entering the buncher is
very low and the bunch-length is undefined.
\beginfig
\endfig
The buncher element therefore has to introduce and (re-) define the bunch
length using the buncher-frequency $f_b$ and the particle velocity $v$.
Hence a \mint-buncher re-samples the longitudinal distribution assuming
a bunch length $\sigma_z$ which is an appropriate fraction of $v/f_b$.
In case of the $870\,\rm{keV}$-beamline, one obtains
\begeq
v/f_b=255\,\rm{mm}
\endeq
\beginfig
\begin{figure*}[t]
\parbox{17.0cm}{
\parbox{12.0cm}{
  \includegraphics[width=120mm]{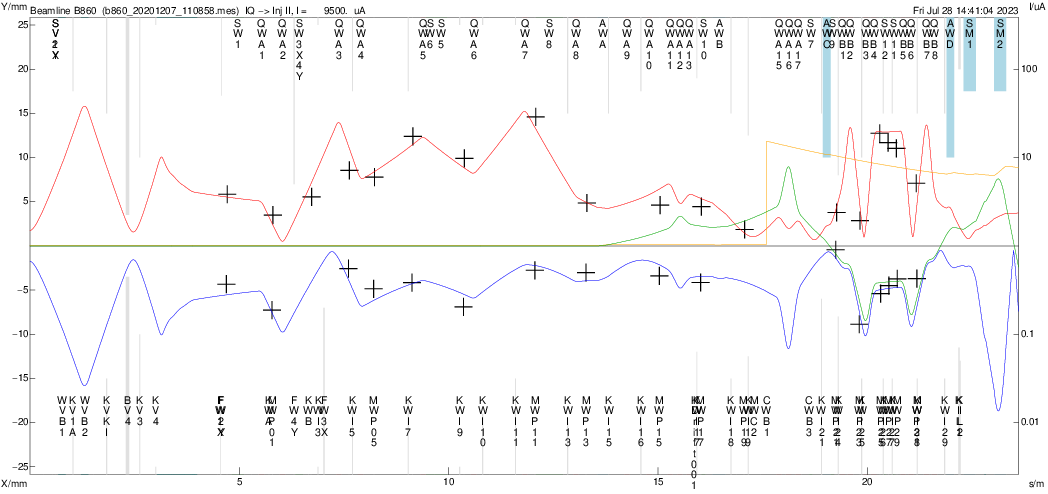}
}\hfill\parbox{4.5cm}{
  \caption[Optics of BW870]{
    Beam envelopes of the BW870 beamline. The bunch length $\sigma_{55}$ is
    shown in orange, but scaled down (here) by a factor $10$. 
    It is undefined (zero) up to the first harmonic buncher CWB.
\label{fig_opt_bw870}}}}
\end{figure*}
\endfig
\mint\ uses typically $2\,\sigma$-values, and hence the bunch length is
assumed to be approximately $4\,\sigma$ in total, so that
\begeq
\sigma_z\approx \frac{1}{4}\,v/f_b\,.
\endeq
\beginfig
\begin{figure*}[t]
\parbox{17.0cm}{
\parbox{8.0cm}{
  \includegraphics[width=80mm]{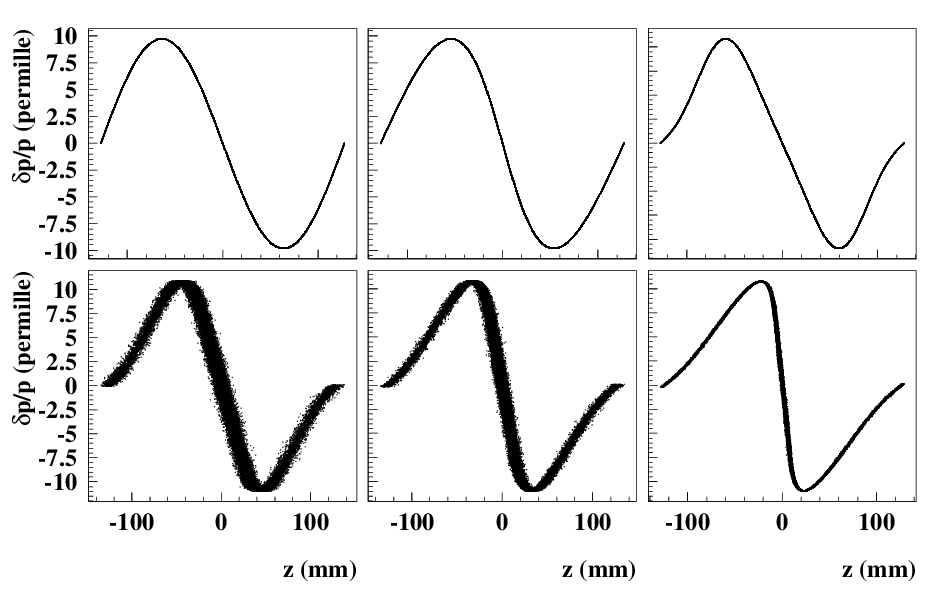}
}\hfill\parbox{8.5cm}{
\caption[Longitudinal Phase Space after Buncher]{
  Evolution of the longitudinal phase space
  of the $870\,\rm{keV}$ injection beamline between from buncher
  (upper left) up to the first turn of Injector 2 (lower right).
  The individual locations are (left to right), $50\,\rm{MHz}$-buncher,
  before and after 3rd harmonic buncher, two positions in the vertical
  injection line (before QWB4 and after QWB6) and finally in the first
  turn of Injector 2. The $50\,\rm{MHz}$-buncher is assumed to have a
  total voltage (both gaps summed up) of $17.1\,\rm{kV}$, the 3rd
  harmonic buncher of $-2.25\,\rm{kV}$.
\label{fig_phasespace}}}}
\end{figure*}
\endfig
After generating a random longitudinal position with this $\sigma_z$,
the energy (i.e. momentum deviation) of each particle is adjusted
accordingly, i.e. the energy change $\Delta E_i$ of the i-th particle
is modified by
\begeq
\Delta E_i=-Q\,V_b\,\sin{(2\,\pi\,z_i/\sigma_z)}\,.
\endeq
where $V_b$ is the buncher voltage. The (linear) evolution of the longitudinal
phase space between buncher and injector is shown in Fig.~\ref{fig_phasespace}.
The core of the formed bunches becomes longitudinally compact, and has a
momentum spread of $\sigma_\delta=0.00682$.
\beginfig
\begin{figure*}[t]
\parbox{17.0cm}{
\parbox{8.5cm}{
  \includegraphics[width=82mm]{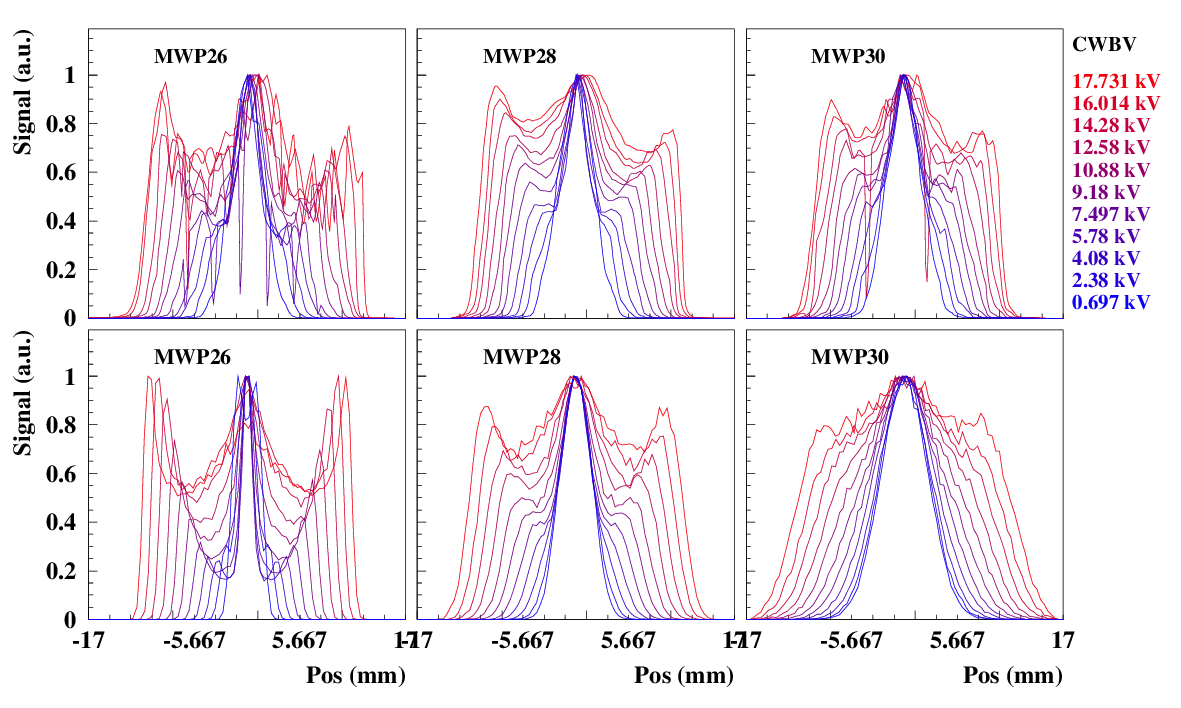}
}\hfill\parbox{8.5cm}{
  \includegraphics[width=80mm]{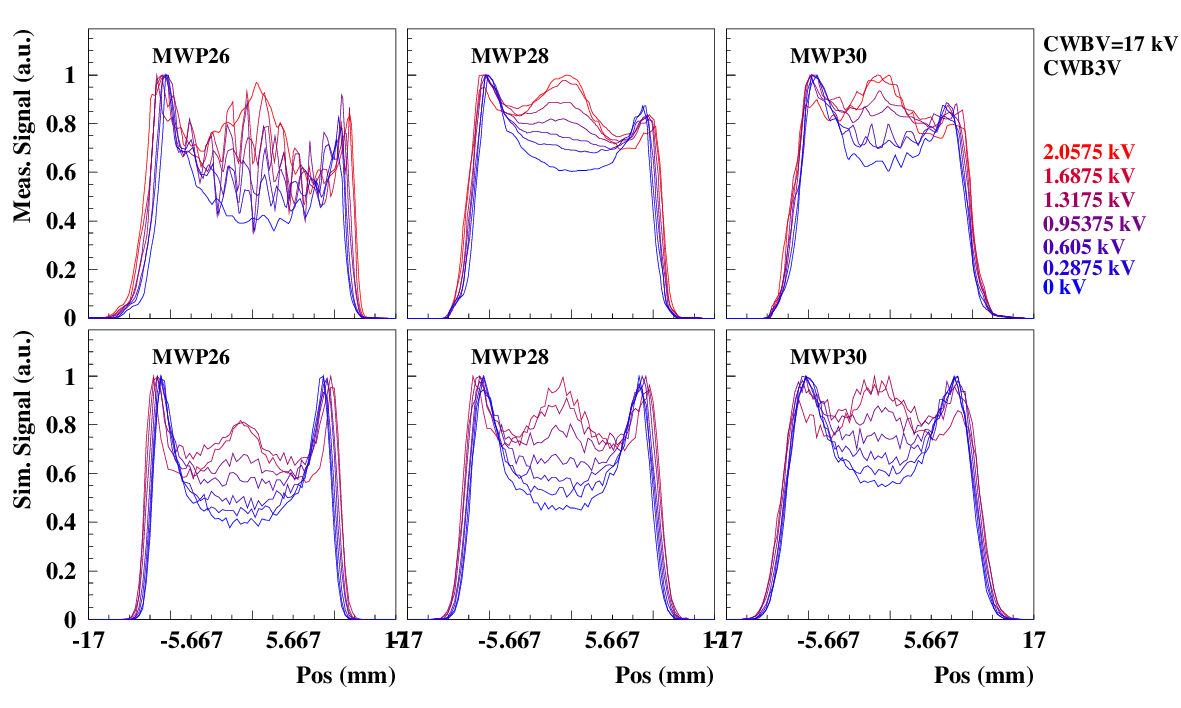}
}}
\caption[Beam Profiles for Different Buncher Voltages]{
    Beam profiles of MWP26, MWP28 and MWP30 in the dispersive section
    of the axial injection line after the bunchers, for various voltages of
    buncher CWB with constant voltage ($2.57\,\rm{kV}$) of the third harmonic
    buncher CWB3 (left). Right: CWB at $17\,\rm{kV}$ with various voltages
    of CWB3. The  measured (simulated) profiles are shown in the top (bottom)
    row. The profiles are centered and normalized to the same maximal value for
    better comparison.
\label{fig_bunchedprofiles}}
\end{figure*}
\endfig
Fig.~\ref{fig_bunchedprofiles} provides a qualitative and quantitative
comparison of the beam profiles in the axial injection line. Since the
dispersion at the location of the profile monitors is non-zero, the
(strongly non-Gaussian) energy distribution induced by the two bunchers
becomes visible in the beam profiles.

There have been simulatios of the bunching process of a DC beamline with 
space-charge presented in the past~\cite{CWB3_07}. The advantage of the
method implemented in \mint\ is that it allows to compare the simulation
results with beam profile measurements, and hence to validate
the used model.

\subsection{The Proton Therapy Beamlines}
\label{sec_proscan}

\begin{figure}[t]
\parbox{17cm}{
\parbox{10cm}{
\includegraphics[width=100mm]{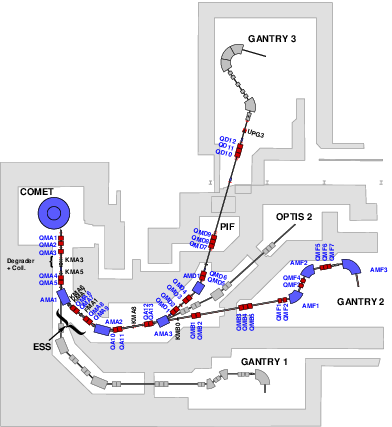}
}\hfill\parbox{6.5cm}{
\caption[Layout of the Proscan facility]{
  Layout of the Proscan facility at PSI with the $250\,\rm{MeV}$
  superconducting cyclotron ``COMET'',
Degrader/Collimator (KMA3 and KMA5), energy selection system ESS, 
Gantry 1~\cite{Pedroni95} (now in the decommissioning-phase), 
Gantry 2~\cite{Pedroni2011}, the eye treatment facility 
OPTIS 2 and the new Gantry 3. The
proton irradiation facility (PIF)~\cite{PIF1} is located 
in the Gantry 3 beam path. Dipole magnets are drawn as blue,
quadrupoles as red boxes. KMA8 and KMB0 are collimators
used for intensity compensation.
\label{fig_zpt_layout}}}}
\end{figure}

Fig.~\ref{fig_zpt_layout} provides an overview of the Proscan
facility~\cite{Pedroni2004,Proscan2012}, where proton beams in an energy
range between $70$ and $230$ MeV are used to irradiate tumors for cancer 
therapy, taking advantage of the so-called bragg peak~\cite{Scharf,ChaoChou,Paganetti,Newhauser}.

The compact isochronous cyclotron ``COMET'' provides a continuous wave (CW)
beam of $250\,\rm{MeV}$ with currents of up to
$800\,\rm{nA}$~\cite{Blosser93,Schillo2001,Geisler2004,Klein2005,Geisler2007,COMET1,COMET2,COMET3}.
The beam energy is adjusted by means of a double-wedge-degrader made
from high-density graphite, followed by a beam collimation system and 
the energy selection system (ESS).
The collimation system consists of two multi-aperture collimators KMA3 and KMA5
and some fixed collimators (KMA4, KMA6 and KMA7)~\cite{Goethem}. A fast kicker 
magnet and beam blocker BMA1 in front of the degrader are used to quickly
switch the beam on or off.

The Proscan facility uses beams in the range from $70\,\rm{MeV}$ to 
about $230\,\rm{MeV}$ for patient treatment, controlled by the degrader wedge 
positions, but never the direct cyclotron beam. The transversal beam emittance 
therefore depends on the collimator geometry -- even at the highest
clinical energy of $230\,\rm{MeV}$ -- but only weakly on the cyclotron beam 
emittance -- as long as the beam is well-focused onto the center of the
degrader wedges.

The (multivariate) Gaussian Monte-Carlo generator implemented in \mint\
approximates the beam distribution from the matrix of second moments by:
\begeq
f(z)={f_{DIS}\over (2\pi)^2\,\sqrt{\vert\Sigma_t\vert}}\,\exp{(-z^T\,\Sigma_t^{-1}\,z/2)}\,,
\endeq
Here only the transversal coordinates are of interest and therefore
$z=(x,x',y,y')$ are the transversal coordinates and $\Sigma_t$ is the
transversal matrix of 2nd moments. But since the collimators KMA3 and KMA5
select the beam in the vicinity of the forward direction, the transmitted 
intensity $I$ passing the collimators on axis (or better: close to the axis) 
can be approximated by:
\myarray{
I&=f(0)\,d\Omega={f_{DIS}\over (2\pi)^2\,\sqrt{\vert\Sigma_t\vert}}\,d\Omega\\
 &={f_{DIS}\over (2\pi)^2\,\eps_x\,\eps_y}\,d\Omega\\
}
where $d\Omega$ is the solid angle of the collimation system and $\eps_i$ are 
the eigenvalues of the $\Sigma_t$-matrix, i.e. the emittances of the
transversal degrees of freedom. $f_{DIS}$ is a beam loss factor that
quantifies beam losses by deep inelastic (large angle-) scattering.

Hence the transmitted (forward) intensity $I$ is approximately inversely 
proportional to the product of the emittances
and hence depends directly on the strength of proton lateral straggling 
inside the degrader. The minimal emittances $\eps_i$ (and hence the maximal
transmission) that can be achieved for a given energy at the degrader exit, 
depends also on the optical properties on the beam entering the 
degrader~\cite{farley1,farley2}, but the emittances of the beam after
the collimation system, is mostly determined by the solid angle defined
by the apertures of the collimation system. 
Hence the beam focus as defined by the quadrupoles QMA1, QMA2 and QMA3
(located in front of the degrader), can be used as an additional knobs to control 
the beam current without having a strong effect on other beam parameters.
TRANSPORT is a powerful tool as long as beam losses and emittance increase along 
the considered beam transport system are negligible. Both of these 
conditions are not met in cyclotron driven proton therapy facilities, where 
energy degraders are used to adjust the beam energy, with the side effect to
increase the beam emittance beyond the acceptance of the beamline. The
use of a degrader implies the necessity to collimate the beam, both
transversally (by beam collimation) and in energy spread by an energy
selection system (ESS).
However this scheme generates a strong energy dependence of the transmitted beam 
intensity. Uncompensated, the beam intensity varies between $70$ and 
$250\,{\rm MeV}$ by roughly three orders of magnitude, for the same 
cyclotron current.
In order to reduce this dynamic intensity range, the beam intensity
change is (partially) ``compensated''~\cite{Pedroni2011}:
At low energies, the beam is well-focused onto the degrader in order to 
provide highest possible transmission. At energies above $\approx
120\,\rm{MeV}$, the beam is intentionally defocused on dedicated collimators
to reduce the energy dependence of the transmitted intensity.

The intensity compensation at Proscan is done in two stages. The first stage
is located upstream of the degrader: QMA3 is used to (de-) focuse the beam
and to smoothly reduce the transmission of high energies.
In order to reduce the dynamic range by two orders of magnitude
(from $\approx 10^3$ to $\approx 10$), a single stage is not sufficient.
A second stage uses the collimator KMA8 for Gantry 3 and, for historical
reasons, another collimator KMB0 for Gantry 2.
Yet again, the beam is focused to achieve high transmission at low energies
and defocused the more the higher the beam energy. Without the second
stage compensation, a single beamline tune (set of magnet settings),
scaled by momentum, would suffice for all energies. The second compensation
stage however requires a slightly different optics setting for each energy
and hence a slightly different optical tune for all energies. 
\beginfig
\begin{figure}[t]
\parbox{17.0cm}{
\includegraphics[width=170mm]{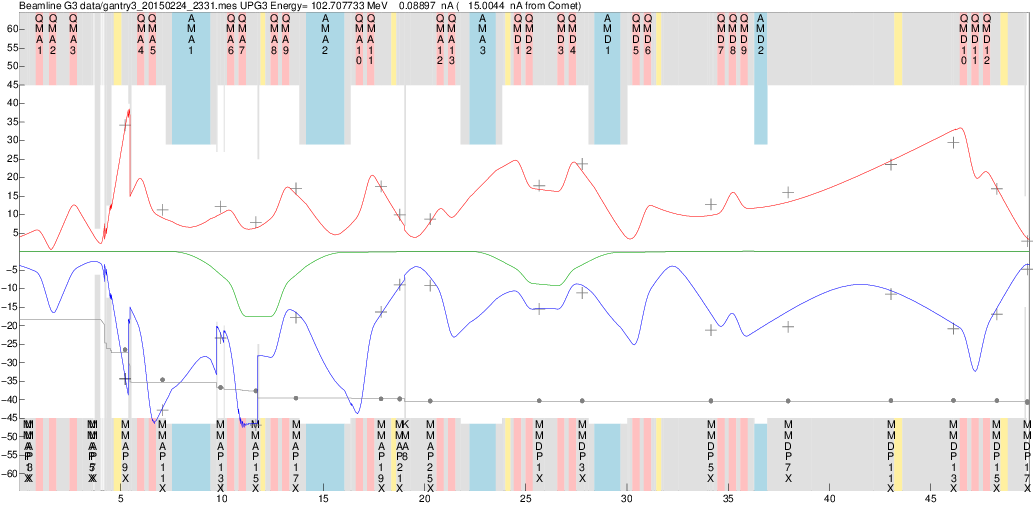}
\caption[Envelope from COMET to Entrance Gantry 3]{
Beam envelopes ($2\,\sigma_{rms}$) of the Proscan beamline
from proton therapy cyclotron COMET to the entrance of 
Gantry 3 for an energy of about $100\,\rm{MeV}$. The (negative) 
dispersion is shown in green. The beam width as measured by the
proscan beam profile monitors~\cite{DiagPT1,DiagPT2} is indicated 
by crosses.
The calculated beam current (i.e. $10\,\log{(I/\rm{\mu A})}$)
is shown as a solid black line (measurements by gray filled
circles). At the end of the beamline, the measured beam current
is $90.4\,\rm{pA}$ agrees well with the \mint\ result $89\,{\rm pA}$.
\label{fig_g3_100}}}
\end{figure}
\endfig

Note also that the proton therapy beamlines use some permanent monitors 
(profile and current) and thin vacuum windows that the beam has to pass~\cite{DiagPT1,DiagPT2}. 
Hence the beam energy after the degrader is not exactly the same as the 
beam energy entering the nozzle. The effects of these monitors/windows 
are small but not completely negligible. The beam optics computation
done with \mint\ allows to take these effects into account.

The patient treatment planning system of Proscan always starts with
the highest required energy (the deepest layer) and the reduces the energy
stepwise for each layer. This is required to avoid hysteresis suppression
cycles between different layers in order to minimize layer switching 
and hence patient treatment time. The beam tunes should therefore preserve 
the ramping direction: a reduction of energy should, for all energies,
correspond to a decrease of the field-settings of all quadrupoles.
Without intensity compensation, the tunes would simply scale the
field with the particle momentum and the requirement would be fulfilled
automatically, but with active intensity compensation this requirement
must be taken into account.

The energy selection system (ESS) consists of a double-bend achromat
composed of two dipoles and four quadrupoles. The first dipole generates
a non-zero dispersion that is compensated by the second dipole.
A moveable collimator (FMA1) is located in the center between the
dipoles where the dispersion is maximal, so that the energy spread of
the beam can be reduced by adjusting the aperture of the horizontal
moveable slit FMA1.

The beamlines to Gantry 1, OPTIS 2 and Gantry 2 are shown 
in gray, as we shall not discuss their optics here.
Gantry 3 shares a beamline with the experimental area of 
the proton irradiation facility (PIF), which is used
only during the night or on weekends, i.e. in times when 
no patient treatments take place. 

\mint\ enables to model the beam tunes of this type of facility,
both qualitatively and quantitatively with reasonable precision. 
The capability of the code to estimate beam-envelopes and -intensity 
simultaneously simplifies the design of beamline layout and tunes 
and allows to predict the transmitted current as well as the locations 
and the amount of beam loss. 
The Monte Carlo mode of \mint\ also enables to predict beam profiles. 
Significant deviations of the predicted beam profiles from measured 
profiles helped in the course of the Gantry 3 project to identify 
errors in the beamline model. Here we show how beam profile and intensity
measurements taken during the commissioning shifts for Gantry 3 compare
to recalculations lately done with \mint.

\beginfig
\begin{figure}[t]
\parbox{17.0cm}{
\includegraphics[width=150mm]{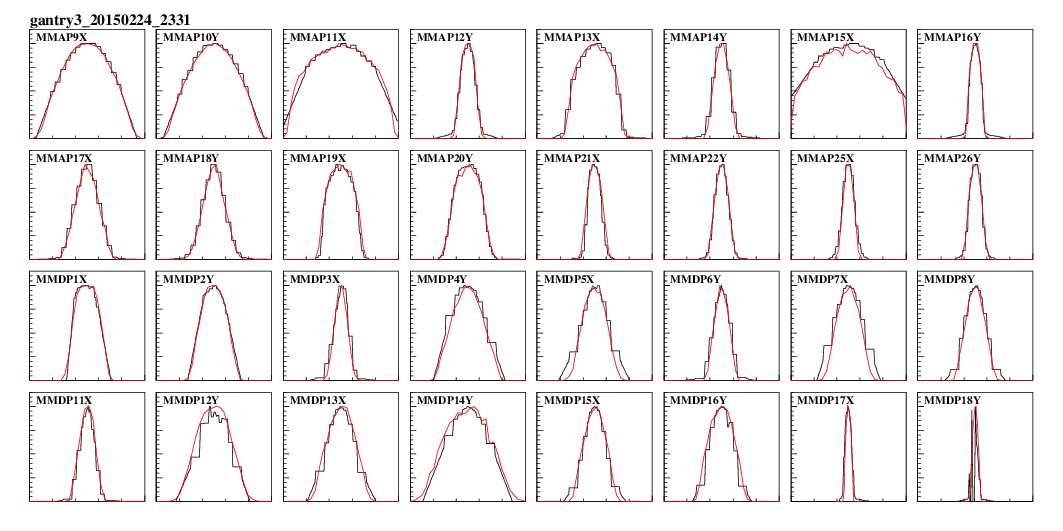}
\caption[Beam Profiles from COMET to Entrance Gantry 3]{
Simulated (red lines) and measured (black lines) beam 
profiles corresponding to the measurement shown in 
Fig.~\ref{fig_g3_100}. The profiles have been equally
scaled and centered for better comparison. (The last
measured profile, MMDP18, is distorted due to a bad 
electrical contact.)
\label{fig_g3_100prof}}}
\end{figure}
\endfig
Fig.~\ref{fig_g3_100} shows a \mint\ calculation of the Gantry 3
beamline up to the coupling point of Gantry 3. This calculation
requires less than $10\,\rm{s}$ with an Monte-Carlo ensemble of
1~Mio. particles on a standard laptop. Simulation results using 
\opal\ of the same beamline have been shown in Ref.~\cite{Rizzoglio}.
\beginfig
\begin{figure}[t]
\parbox{17.0cm}{
\includegraphics[width=150mm]{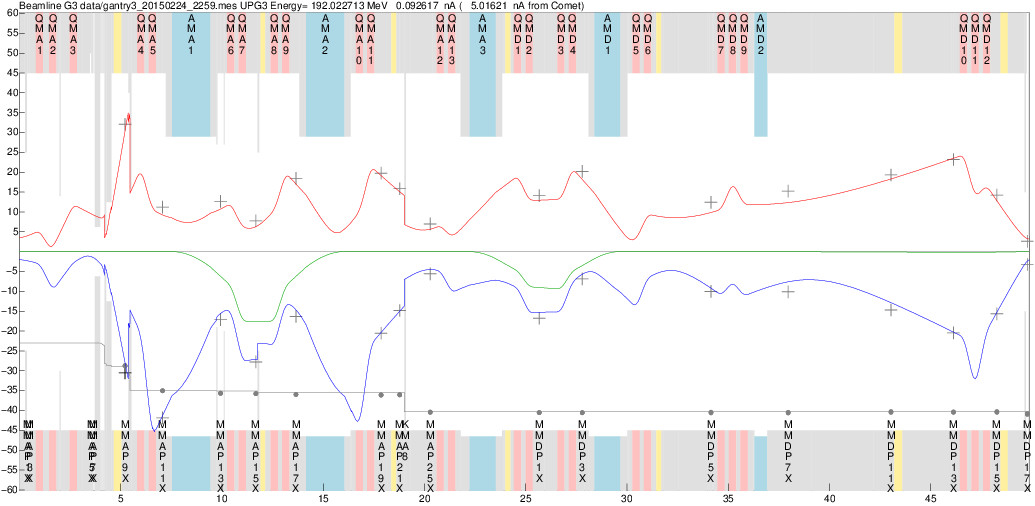}
\caption[Envelope from COMET to Entrance Gantry 3 (190 MeV)]{
Same plot as is Fig.~\ref{fig_g3_100}, but for an energy of
$\approx 190\,\rm{MeV}$. The intensity is reduced by defocusing
the beam vertically with quadrupole QMA3 and after the ESS
on collimator KMA8. The Monte Carlo mode starts after the
degrader such that the collimator in front of the degrader
(after QMA3) has no effect, whereas the collimators behind 
the degrader (at about $s=4\,\rm{m}$) and collimator KMA8 
(at about $s=18\,\rm{m}$) reduce both, beam size and current,
in agreement with the measurements.
\label{fig_g3_190}}}
\end{figure}
\endfig

\section{Summary and Outlook}

A description of the methods used in the new ion beam optics program \mint\
has been given, which allow to extend the applicability of TRANSPORT type
calculations to include the passage of matter and beam collimation in a fast
and effective way.

The results of various ion beam optics calculations with \mint have been 
compared with beam profile and intensity measurements of the PSI proton
facilities, firstly the high intensity facility HIPA, and secondly of the
proton therapy facility Proscan. In all cases \mint\ provides convenient
and fast methods to simulate beam line optics and to compare model and
measurement. 
The accuracy of the ion optics calculation is comparable to TRANSPORT's first
order calculations but the range of applicability has been expanded.
\mint\ allows additionally for the simulation of slices of energy degraders
or targets and -- in the Monte-Carlo-mode -- of collimators with simple
geometry. It can predict beam losses and realistic beam profiles.
The accuracy of these latter calculations is of course limited by the
precision of the various multiple scattering approximations. The best
agreement with data of the proton therapy facility was obtained using
Gottschalk's scattering model~\cite{Gottschalk2}.

\mint\ also allows for the calculation of matched beams in ion (storage)
rings, cyclotrons, and FFA's for arbitrary couplings and given emittances including 
space-charge kicks. \mint\ also allows for the iterative calculation of linear
matched beams, even for cyclotron specific couplings, i.e. the so-called ``vortex
effect''. The correspondings phase space ensembles can be used as starting
conditions in other codes like \opal~\cite{cyc_paper,cyc2013,cyc2019}.

\mint\ does not aim to compete with more ``realistic'' codes like \opal\ or
BDSIM. \mint\ is intended to replace (and extend the capabilities of) the fast
and lightweight codes TRANSPORT and TURTLE, specifically as a control room
tool for beamlines like those of PSI's proton accelerator facilities.
\mint\ has not been (directly) validated by comparison with other codes,
but by comparison with profile and current measurements performed at PSI.

\section{Acknowledgements}

We thank Hubert Lutz and Jochem Snuverink for their kind support in
installing \mint\ on the control system computers and Corina Sattler
for providing Fig.~\ref{fig_hipa}.
The other figures have been generated with \mint\ or with the cernlib
(PAW) and XFig, respectively. 

\mint\ has been written in ``C++'' and compiled with the GNU\textsuperscript{\copyright}-C++
compiler on various Linux systems (Scientific Linux, Red Hat 6 \& 7 \& 8, OpenSuSE 15.X,
Ubuntu). \mint\ uses the GNU\textsuperscript{\copyright}
scientific library (GSL), and the GNU programs flex and bison. 
\mint\ version 0.50, described here, uses GNU libplot for 
graphical output. 

\begin{appendix}

\section{Some Features of \mint}

\subsection{The Unit System of \mint}

The main purpose of \mint\ is to provide an alternative to
TRANSPORT for the PSI beamlines, both as an online-tool for 
the machine control rooms but also as a tool to develop tunes 
for Proscan or layouts for new beamlines.

\mint\ offers (and to a large degree requires) the definition and 
the use of units. The predefined unit system is the SI-system,
but the user is free to define and use other units, for instance
imperial units. \mint\ automatically scales variables accordingly
and checks the consistency of calculations with respect to their
physical units:
\mint\ allows to add/subtract or assign only quantities of the
same physical dimension. The user not only has the option but
is obliged to define the units of all inputs. But he enjoys the
freedom to define all parameters in his preferred units and there
is no need for the user to convert units ``manually''.

\mint\ comes with a considerable number of pre-defined (SI-) units.
Also many ``physical constants'' are predefined in the form of physical
units, for instance the speed of light 'c', the elementary charge 'e',
the mass of the electron 'Me' and proton 'Mp' and many more. This
allows to formulate physical equations in a convenient and
readable manner. The \mint\ syntax expects units written
in single quotes in order to distinguish, for instance, 'm/s'
from a quotient of variables.

Since certain variables, like for instance the pole tip field of 
a quadrupole magnet, make no sense unless they have the
correct unit, \mint\ allows the user to define variables with
a fixed physical meaning, like ``MagneticField B'' (instead of ``Var B'').
That is, the type of the physical quantity to be represented by
a variable, can be generically taken into account. Hence, strictly
speaking, \mint\ has at least as many basic types as there are
different physical quantities and the unit system can be understood
as a type checker. 

\subsection{The Class/Type System of \mint}

The \mint\ input is based on a simple programming language but
provides some features of object-oriented programming. There
are predefined ``objects'', namely beamline elements like drifts,
bends and quads, but additional data fields can be defined and part of 
their functional behavior can be modified and some basic typing
with feature inheritance has been implemented. \mint\ allows 
to create hierarchies of element types by the use of
user-defined beamline element classes. If values are specified (assigned) 
within the type definition, then they are properties of the class and 
hence, if changed, are changed in all instances of that class. 
This is due to the fact that variables which are initialized within 
a type-definition, appear only once in memory and are therefore
valid for all instances of the respective type. 
{\scriptsize\begin{verbatim}
Type MyQuad(Quad) {
   SHAPE = CIRCULAR;
   Options = OPTLABEL;
   L = 368.0 'mm';
   DS = L/4.0;    // Step size
   R = 50.0 'mm'; // Pole tip radius
   RX = R-5.0 'mm'; // Hor. beam pipe radius
   RY = R-5.0 'mm'; // Vert. beam pipe radius
};

// Since ``Current'' and ``MagneticField'' are predefined physical
// quantities, they can be used as if they were basic data types:
Type QMA(MyQuad) {
   Current I,Ilin=101.8 'A',Imax=150.0 'A';
   MagneticField b0 = 34.8 'G' , b1 = 8856.5 'G', 
     b2 = -671.4 'G' , b3 = -80.5 'G';
   x:=(abs(I)-Ilin)/(Imax-Ilin);
   B:= sign(I)*(b0 + b1 * abs(I)/Imax + 
     (b2 * x^2 + b3 * x^3)*theta_h(abs(I)-Ilin));
};

[...]
Beamline G3 { 
[...]
  QMA QMA1 { I= 107.35 'A'; };
};
[...]
\end{verbatim}}
The length of the quadrupole ``QMA'' is, as defined here, a property 
of the class ``MyQuad''. If this length is redefined later, this changes
the length of all quads of this {\it class} and hence the length of all 
quadrupole instances of this type. This allows to fit the
parameters not only of individuals but of types; however it requires
some care in the definition of the type hierarchy. 

The actual definition of quadrupole ``QMA1'' of type ``QMA'' requires  
only to define the coil current to be completed. Since the field value
``B'' has been (re-) defined, in this example, as a function 
(indicated by ``:=''), it is evaluated on every read access~\footnote{
In the current version of \mint, this is the only way for the user to 
define functions.}.
When the ion optics machinery of \mint\ reads the quadrupole field in 
order to compute the transfer matrix, the function ``x'' is evaluated 
and hence the field value is effectively a function of the (user-defined)
current-variable ``I''.

The excitation curve is approximated here by a third order polynomial
and from this, the pole tip field is 
automatically computed, whenever it is used. Since \mint\ computes the
transfer matrix of each element just before the matrix is to be used,
one might as well define:
{\scriptsize\begin{verbatim}
Type QMA(Quad) {
   [...]
   MagneticField B0;
   B:=B0*(PC/('Mp' * 'c' * 'c'));
};
[...]
Beamline G3 { 
[...]
  QMA QMA1 { B0 = 7.23 'kG'; };
};
\end{verbatim}}
where ``PC'' is the predefined variable of the beam's actual momentum
and ``'Mp''' the projectile (here: proton) mass, 
so that now the quad field strength is automatically scaled to the
momentum and equals ``B0'' if the momentum (times the speed of light)
is equal to $m\,c^2$ (that is, for $\y=\sqrt{2}$ or $\beta=1/\sqrt{2}$).
Since \mint\ allows for stacked include statements, the user
can build up type-libraries for specific kind of problems and easily 
switch between them by modifying merely the include-statements.

The input script is, like in high-level languages, devided into two parts. 
The first part provides the definition of element type and a description 
of the beamline(s). The main part is the ``program'', i.e.
contains the operations to be performed with the beamline. 
The main part of a mint-script is translated into a sequence 
of ``byte-codes'' which then allows to implement some standard 
loops (for-do, if-then-else, while-do, repeat-until).
The overhead of generating a byte-code and runtime-linking does
not significantly increase the execution time, since most of
the execution time is usually spend with matrix multiplications.

In order to better compare results with TRANSPORT, \mint\ provides
the possibility to export beamlines to a TRANSPORT type input
file. However, due to the different concepts of the two programs,
some minor adjustments by hand are usually required before the
use with TRANSPORT~\footnote{For instance, TRANSPORT does not
accept layout calculations combined with space charge. The
user then has to actively deactivate either of the two before
running TRANSPORT.}. 

\subsection{Misalignments in \mint}

\mint\ allows to define a number of misalignments as for instance
offsets ``Xofs'' (horizontal position error) and ``Yofs'' (vertical position
error) and misalignments of the axis {\it Pitch, Yaw and Roll}.
If these variables are not initialized by the user, they are assumed to
be zero. Trivially \mint\ can, beyond the envelope calculation, calculate,
plot and fit the beam centroid as well. These features have been used to
determine some (small) quadrupole misalignments within the Proscan facility.
Details shall be reported elsewhere.

\subsection{Fitting with \mint}

The fitting and optimization functions of \mint\ are based upon
the GNU scientific libraries (GSL) minimization routines. Each
beamline-element in \mint\ is equipped with a {\it malus}-function
which is supposed to quantify the disagreement between the actual
state and the desired state of the beamline. Some elements,
specifically monitors, have a predefined malus-calculation which
computes the (squared) deviation of the calculated and measured
beam size, which is activated whenever a non-zero measurement
of the beamsize is defined by the user. The malus-functions of
other beamline elements must be defined by the user according to
their needs.

A typical malus-definition within an element-declaration
might be~\footnote{{\it maxError} and
  {\it minError} are predefined \mint-functions which compute the
  conditioned squared deviation: The result of {\it maxError} is zero
  if the respective value is below the limit.}:
\begin{verbatim}
Malus:=maxError(s11,RX/2.0,0.25 'mm')+maxError(s33,RY/2.0,0.25 'mm');
\end{verbatim}
which increases the {\it malus} in case of a too large beam size, i.e.
in cases where $\sigma_x=s11$ and/or $\sigma_y=s33$ are larger than
half of the user-defined radius $R_x$ and $R_y$. The precision is also
defined (here $0.25\,\rm{mm}$). \mint\ profile monitors set the
respective {\it malus}-variables to the square deviation, provided
that the measured beam size and centroid positions are non-zero.

The {\it malus} of the beamline is the sum of the {\it malus} variables
of all elements and the miminization of the {\it malus} is done by the
beamline-method {\it Vary()}:
\begin{verbatim}
PKSINQ::Vary(@PKSINQ.Malus,@PKSINQ.QHG21.I1,@PKSINQ.QHG21.I2,@PKSINQ.QHG21.I3);
\end{verbatim}
where the first argument refers to the {\it malus}-variable to be minimized
and all following parameters are to be varied (the number of parameters to
be varied is arbitrary).

\mint\ is also equipped with an extension of this fitting routine which allows
to vary multiply settings ``at once''. This is the {\it MultiVary}-method, in
which the {\it malus} is summed over a set of $N$ calculations before the
parameters are varied. This feature can for instance be used, if a parameter
can not be fitted based on a single beamline setting. The syntax is:
\begin{verbatim}
G3::MultiVary(@iTune,17,@G3.Malus,@G3.Protons.s11,@G3.Protons.c12,@G3.Protons.s22);
\end{verbatim}
where the first parameter, {\it iTune}, is an integer variable that indicates
the index of the ``run'', the second parameter gives the number of ``runs''
and the remaining parameters are equivalent to {\it Vary()}, i.e. the {\it
  malus}-variable and the parameters to be varied. In the given example, the
initial beamsize (of the initial beam named {\it protons}) is varied using
$17$ runs. The user then must take care that the respective beamline
parameters are properly selected by the iterator ``iTune''. Since
\mint\ allows to define {\it vectors}, {\it matrices} and (simple) functions,
the user can use a construction like the following:
{\scriptsize\begin{verbatim}
Int iTune=0;
Vector qma_fields[4]={ 3.7 'kG', 4.711 'kG', 5.1 'kG', 0.6123 'T' };

Type QMA(Quad) {
  [...]
  B:=qma_fields[iTune];
};

Beamline BL {
  ...
  QMA QMA1;
  ...
};

begin
  BL::MultiVary(@iTune,4,@BL.Malus,@BL.SomeElement.SomeVariable,...);
end.
\end{verbatim}
}
This will cause \mint\ to sum up the {\it malus}-variable four times (starting
from the initial value) each time, before \verb!@BL.SomeElement.SomeVariable! is varied
to minimize the {\it malus}. This is useful to fit for instance misalignment
variables like quadrupole offsets to reproduce {\it sets} of position
measurements for {\it sets} of quadrupole settings (``beam tunes'').

\subsection{Multiple Plots in \mint}

\mint\ allows to plot a number of runs on a single graph.
\beginfig
\begin{figure}[t]
\parbox{17.0cm}{
\includegraphics[width=150mm]{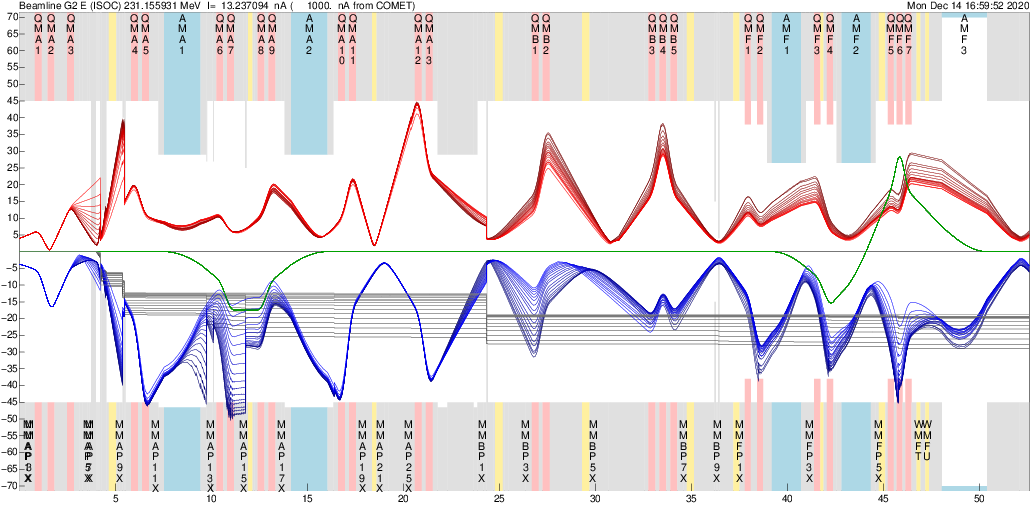}
\caption[Envelopes of Tunes from COMET exit to Gantry 2 Isocenter]{
  TRANSPORT style plot of the beam envelopes for a set of tunes
  for Gantry 2, starting from the cyclotron COMET (left to right)
  up to the isocenter. The second compensation collimator ``KMB0''
  is located at $\approx 24.5\,\rm{m}$ downstream of profile monitor
  ``MMBP1X''. Note that the Monte-Carlo mode is here activated
  {\it behind} the degrader so that the collimators {\it before}
  the degrader are ignored (behind QMA3). This is done on purpose
  to reduce computation time.
\label{fig_g2tunes}}}
\end{figure}
\endfig
Fig.~\ref{fig_g2tunes} shows for example $17$ optimized tunes for
the Gantry-2 beamline of Proscan covering the range from $70\,\rm{MeV}$
to $230\,\rm{MeV}$ in steps of about $10\,\rm{MeV}$ thus illustrating
the use of the intensity compensation scheme by the use of two collimators
located after QMA3 and QMA13, respectively. The energy dependence of
the horizontal energy spread and hence the horizontal beam size
in the dispersive region between AMA1 and AMA2 is nicely visualized
in this plot.

The methods used to do this are:
\begin{verbatim}
for (k=0;k<17;k+=1) {
  [...]
  G2::Envelope("g2_optics_%d.env"<k);
  G2::PushOptics();
  [...]
};
[...]
G2::FlushOptics("X","",1200,600);
G2::FlushOptics("PS","g2_kmb0_tunes_v1.ps",12000,6000);
G2::ClearOptics();
\end{verbatim}

\subsection{Layout Calculations and Plots in \mint}

\begin{figure}[t]
\parbox{17.0cm}{
\parbox{11.0cm}{
  \includegraphics[width=105mm]{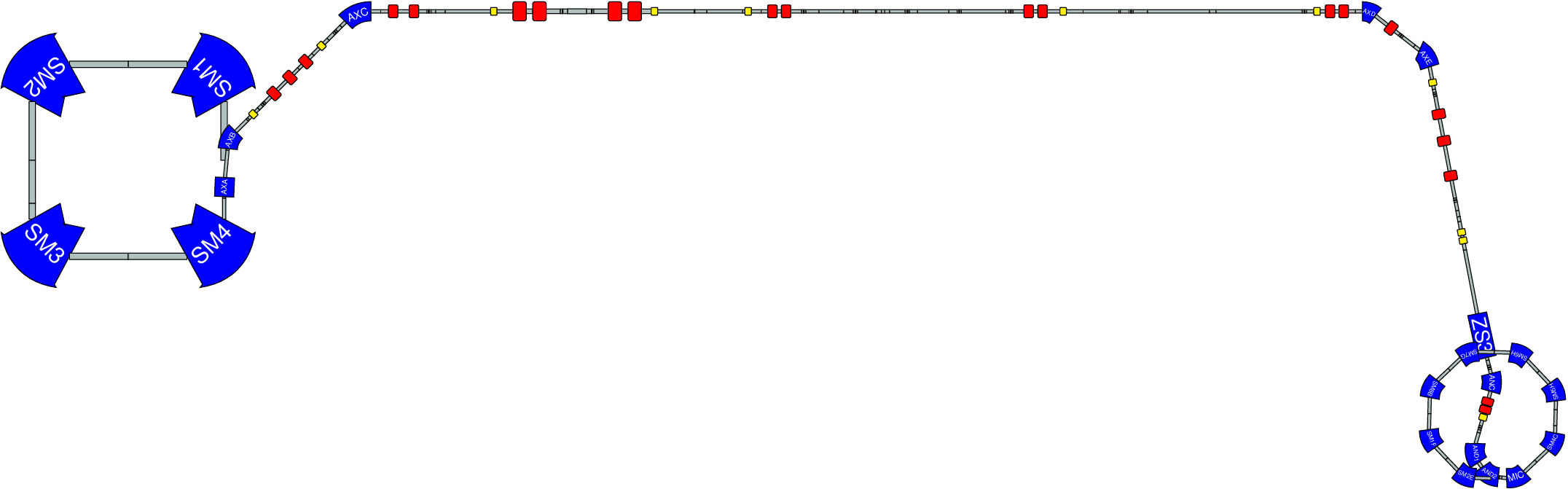}
}\hfill\parbox{6cm}{
\caption[Layout of IW2]{
  Layout of the beamline IW2, including the last turn of Injector 2
  and the first turn of the Ring cyclotron. (The corresponding
  beam optics is shown in Fig.~\ref{fig_iw2_fit}). This plot
  was generated from a FIG-file (for the Linux program XFig)
  produced by \mint, but a direct Postscript output is possible
  as well.
\label{fig_iw2_layout}}}}
\end{figure}

\mint\ allows to compute the floor layout of beamlines and to
produce figures illustrating the result. Fig.~\ref{fig_iw2_layout}
shows the graphical output produced by \mint-layout routines. 
\beginfig
\endfig
The main purpose of the Layout routines, however, is not to generate
beautiful graphics, but to provide numerical layout data to be compared
survey data. An example of the output format:
{\small\begin{verbatim}
   Element    Type        L          S_ref    Z_vertex    Z_traj   X          Y          Z
   -           -          'm'        'm'       'm'        'm'     'm'        'm'        'm'
       Sec       ISEC   5.494782    0.12875         0.         0.         0.         0.         0.
  Valley2b         ZS   1.834156   5.623532   6.540384    6.54061 568.198999 216.619893        1.5
       SM2         SM    1.82647   7.457688   8.620229   8.370923 566.119154 216.619893        1.5
  Valley3a         ZS   1.834156   9.284158  10.700074  10.201236 566.119154 214.540048        1.5
 Valley3b1      Drift      1.351  11.118314  12.292652  11.793814 566.119154  212.94747        1.5
       RIZ      ProfZ         0.  12.469314  12.968152  12.469314 566.119154  212.27197        1.5
 Valley3b2      Drift    0.48778  12.469314  13.212042  12.713204 566.119154  212.02808        1.5
       SM3         SM    1.82647  12.957094  14.618699  13.870329 566.119154 210.621423        1.5
  Valley4a         ZS   1.834156  14.783564  16.698543  15.700642 568.198999 210.621423        1.5
  Valley4b         ZS   1.834156   16.61772  18.532699  17.534798 570.033155 210.621423        1.5
       SM4         SM    1.82647  18.451876  20.612544  19.365111    572.113 210.621423        1.5
      MXP1    ProfX20         0.  20.928346  22.425311  20.928346    572.113  212.43419        1.5
       AXA      ABend        0.6  20.978346  22.775534  21.278346    572.113 212.784412        1.5
       FXE     RDrift         0.  22.431346  23.928756  22.431346 572.221528 213.932516        1.5
       AXB      ABend      0.602  22.481346  24.292339  22.782346 572.255744 214.294486        1.5
       ...
\end{verbatim}}
The output data are aligned with the PSI convention, according to which
there are two positions along the beam line. $S_{ref}$ is the measured
distance along the actual trajectory and follows the actual path in dipole
magnets. $Z_{traj}$ is the length of the beam path if described as a polygon
connecting the vertices of the bending magnets. It is little more than
a convenient measure to compare results with geometric layout data and
technical drawings.

\section{Quadrupole Fringe Fields}
\label{sec_qfringes}

\subsection{TRANSPORT method (Matsuda/Wollnik)}

Note that \mint\ provides only the linear approximation of the method
proposed by Matsuda and Wollnik~\cite{Matsuda}.
\beginfig
\begin{figure}[t]
\parbox{17.0cm}{
\parbox{13.0cm}{
  \includegraphics[width=125mm]{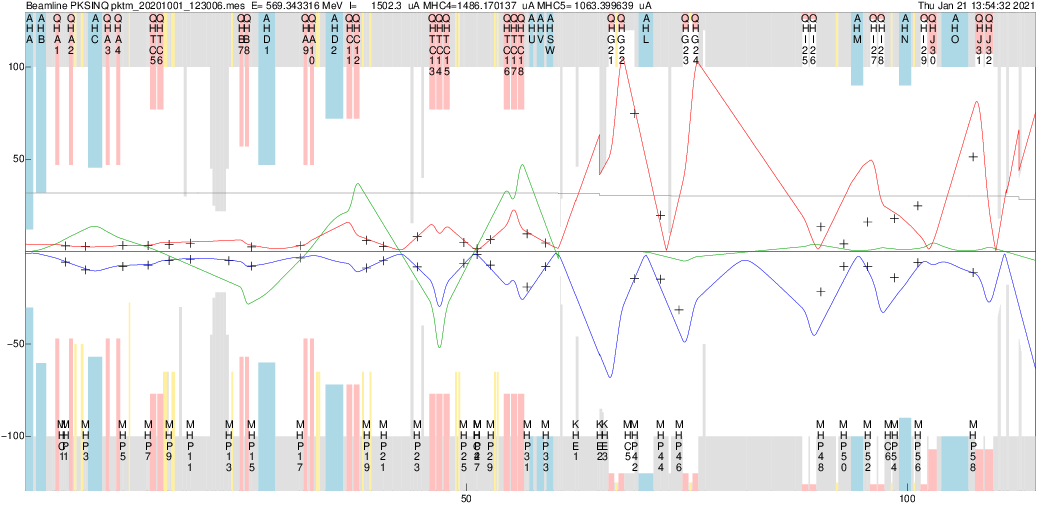}
}\hfill\parbox{4cm}{
\caption[]{
Computed beam envelope as in Fig.~\ref{fig_pksinq40}, but 
here without any quadrupole fringe field corrections.
The measured beam sizes (symbols) can not be reproduced
without a reasonable fringe field correction for
large aperture quadrupoles.
\label{fig_pksinq40_nofringes}}}}
\end{figure}
\endfig
Provided the fringe field integrals $I_1$, $I_2$ and $I_3$ are
known according to Rohrer's TRANSPORT convention, the single
quadrupole matrix ${\bf M}_q$ is embraced by two additional
matrices at the quadrupole entrance ${\bf M}_i$ and exit ${\bf M}_f$:
\begeq
{\bf M}_q\to{\bf M}_f\,{\bf M}_q\,{\bf M}_i\,.
\endeq
The horizontal ${\bf M}_h$ and vertical ${\bf M}_v$ $2\times 2$-matrices are given by:
\myarray{
{\bf M}_h&=\frac{1}{a}\,\bmtx{cc}
1-x_0 & -x_1\\
-x_2 & 1+x_0\\
\emtx\\
{\bf M}_v&=\frac{1}{b}\,\bmtx{cc}
1+x_0 & x_1\\
-x_2 & 1-x_0\\
\emtx
}
where (upper sign for entrance, lower for exit): 
\myarray{
  x_0&=\pm\,K\,R^2\,I_1\\
  x_1&=2\,K\,R^3\,I_2\\
  x_2&=K^2\,R^3\,I_3\\
}
where $K$ is the quadupole strength $K={B\over R\,(B\rho)}$,
which is positive for horinzontally focusing quads.
The variables $a$ and $b$ are both equal to $1$ in the
original calculation. However, $2\times 2$-matrices are
only symplectic, if they have unit determinant. Hence
\mint\ uses:
\myarray{
  a&=\sqrt{1-x_0^2-x_1\,x_2}\\
  b&=\sqrt{1-x_0^2+x_1\,x_2}\\
}

\subsection{Baartman's approach to quadrupole fringe fields}

Baartman's correction~\cite{Baartman} is implemented by defining the following 
variables:
\begary{rclp{10mm}rcl}
\Delta\alpha_f&=&f_0\,k^2\,R^2/2\,{s^2\over s+\alpha\,c}&&k_f&=&k\,(1-{\Delta\alpha_f\over\alpha})\\
\Delta\alpha_d&=&f_0\,k^2\,R^2/2\,{S^2\over S+\alpha\,C}&&k_d&=&k\,(1+{\Delta\alpha_d\over\alpha})\\
\endary
and from these new values the corrected solutions:  
\begary{rclp{10mm}rcl}
\alpha&=&k\,L&& &&\\
c'&=&\cos{(\alpha-\Delta\alpha_f)}&&s'&=&\sin{(\alpha-\Delta\alpha_f)}\\
C'&=&\cosh{(\alpha+\Delta\alpha_d)}&&S'&=&\sinh{(\alpha+\Delta\alpha_d)}\\
\endary
and the corrected (focusing/defocusing) sub-matrices:
\begary{rcl}
{\bf M}_f&=&\bmtx{cc}
c'&s'/k_f\\
-k_f\,'s&c'\\
\emtx\\
{\bf M}_d&=&\bmtx{cc}
C'&S'/k_d\\
k_d\,S'&C'\\
\emtx
\endary
The results of this method for the SINQ-beamline are shown in Fig.~\ref{fig_pksinq40}.
If quadrupole fringe fields are ignored, the optics can not be reproduced
(See Fig.~\ref{fig_pksinq40_nofringes}).

\end{appendix}

\bibliography{mint_paper}{}
\bibliographystyle{unsrt}

\end{document}